\documentclass[fleqn, usenatbib, useAMS]{mnras}

\usepackage{newtxtext,newtxmath}
\usepackage[T1]{fontenc}
\usepackage{ae,aecompl}
\usepackage[utf8]{inputenc}

\usepackage{graphicx}	
\usepackage{mathtools}
\usepackage{bm}		    

\newcommand{\fullderiv}[2]{\frac{d #1}{d #2}}

\newcommand{\Tr}{\mathrm{Tr}}

\newcommand{\skm}{~s~km$^{-1}$\;}
\newcommand{\kms}{~km~s$^{-1}$\;}

\newcommand{\poned}{$P_{\mathrm{1D}}$\;}
\newcommand{\ddif}{\mathrm{d}}

\title[Framework to measure metal properties]{A framework to measure the properties of intergalactic metal systems with two-point flux statistics} 

\author[N. G. Kara{\c c}ayl{\i} et al.]{
\parbox{\textwidth}{
\Large
Naim G{\" o}ksel Kara{\c c}ayl{\i}$^{1, 2, 3}$,
Paul Martini$^{1, 2}$,
David H. Weinberg$^{2}$,
Vid Ir{\v s}i{\v c}$^{4}$,
J.~Aguilar$^{5}$,
S.~Ahlen$^{6}$,
D.~Brooks$^{7}$,
A.~de la Macorra$^{8}$,
A.~Font-Ribera$^{9}$,
S.~Gontcho A Gontcho$^{5}$,
J.~Guy$^{5}$,
T.~Kisner$^{5}$,
R.~Miquel$^{9,10}$,
C.~Poppett$^{5,11,12}$,
C.~Ravoux$^{13}$,
M.~Schubnell$^{14}$,
G.~Tarl\'{e}$^{14}$,
B.~A.~Weaver$^{15}$,
and Z.~Zhou$^{16}$
}
\vspace{0.4cm}
\\
\parbox{\textwidth}{
$^{1}$ Center for Cosmology and AstroParticle Physics, The Ohio State University, 191 West Woodruff Avenue, Columbus, OH 43210, USA\\
$^{2}$ Department of Astronomy, The Ohio State University, 4055 McPherson Laboratory, 140 W 18th Avenue, Columbus, OH 43210, USA\\
$^{3}$ Department of Physics, The Ohio State University, 191 West Woodruff Avenue, Columbus, OH 43210, USA\\
$^{4}$ Kavli Institute for Cosmology, University of Cambridge, Madingley Road, Cambridge CB3 0HA, UK\\
$^{5}$ Lawrence Berkeley National Laboratory, 1 Cyclotron Road, Berkeley, CA 94720, USA\\
$^{6}$ Physics Dept., Boston University, 590 Commonwealth Avenue, Boston, MA 02215, USA\\
$^{7}$ Department of Physics \& Astronomy, University College London, Gower Street, London, WC1E 6BT, UK\\
$^{8}$ Instituto de F\'{\i}sica, Universidad Nacional Aut\'{o}noma de M\'{e}xico,  Cd. de M\'{e}xico  C.P. 04510,  M\'{e}xico\\
$^{9}$ Institut de F\'{i}sica d'Altes Energies (IFAE), The Barcelona Institute of Science and Technology, Campus UAB, 08193 Bellaterra Barcelona, Spain\\
$^{10}$ Instituci\'{o} Catalana de Recerca i Estudis Avan\c{c}ats, Passeig de Llu\'{\i}s Companys, 23, 08010 Barcelona, Spain\\
$^{11}$ Space Sciences Laboratory, University of California, Berkeley, 7 Gauss Way, Berkeley, CA  94720, USA\\
$^{12}$ University of California, Berkeley, 110 Sproul Hall \#5800 Berkeley, CA 94720, USA\\
$^{13}$ IRFU, CEA, Universit\'{e} Paris-Saclay, F-91191 Gif-sur-Yvette, France\\
$^{14}$ Department of Physics, University of Michigan, Ann Arbor, MI 48109, USA\\
$^{15}$ NSF's NOIRLab, 950 N. Cherry Ave., Tucson, AZ 85719, USA\\
$^{16}$ National Astronomical Observatories, Chinese Academy of Sciences, A20 Datun Rd., Chaoyang District, Beijing, 100012, P.R. China\\
}
}

\date{Accepted XXX. Received YYY; in original form ZZZ}

\pubyear{2022}

\begin{document}
\label{firstpage}
\pagerange{\pageref{firstpage}--\pageref{lastpage}}
\maketitle

\begin{abstract}
The abundance, temperature, and clustering of metals in the intergalactic medium are important parameters for understanding their cosmic evolution and quantifying their impact on cosmological analysis with the Ly~$\alpha$ forest. The properties of these systems are typically measured from individual quasar spectra redward of the quasar's Ly~$\alpha$ emission line, yet that approach may provide biased results due to selection effects. We present an alternative approach to measure these properties in an unbiased manner with the two-point statistics commonly employed to quantify large-scale structure. Our model treats the observed flux of a large sample of quasar spectra as a continuous field and describes the one-dimensional, two-point statistics of this field with three parameters per ion: the abundance (column density distribution), temperature (Doppler parameter) and clustering (cloud-cloud correlation function). We demonstrate this approach on multiple ions (e.g., \ion{C}{IV}, \ion{Si}{IV}, \ion{Mg}{II}) with early data from the Dark Energy Spectroscopic Instrument (DESI) and high-resolution spectra from the literature. Our initial results show some evidence that the \ion{C}{IV} abundance is higher than previous measurements and evidence for abundance evolution over time. The first full year of DESI observations will have over an order of magnitude more quasar spectra than this study. In a future paper we will use those data to measure the growth of clustering and its impact on the Ly~$\alpha$ forest, as well as test other DESI analysis infrastructure such as the pipeline noise estimates and the resolution matrix.
\end{abstract}




\section{Introduction}
\begin{figure*}
    \centering
    \includegraphics[width=0.8\linewidth]{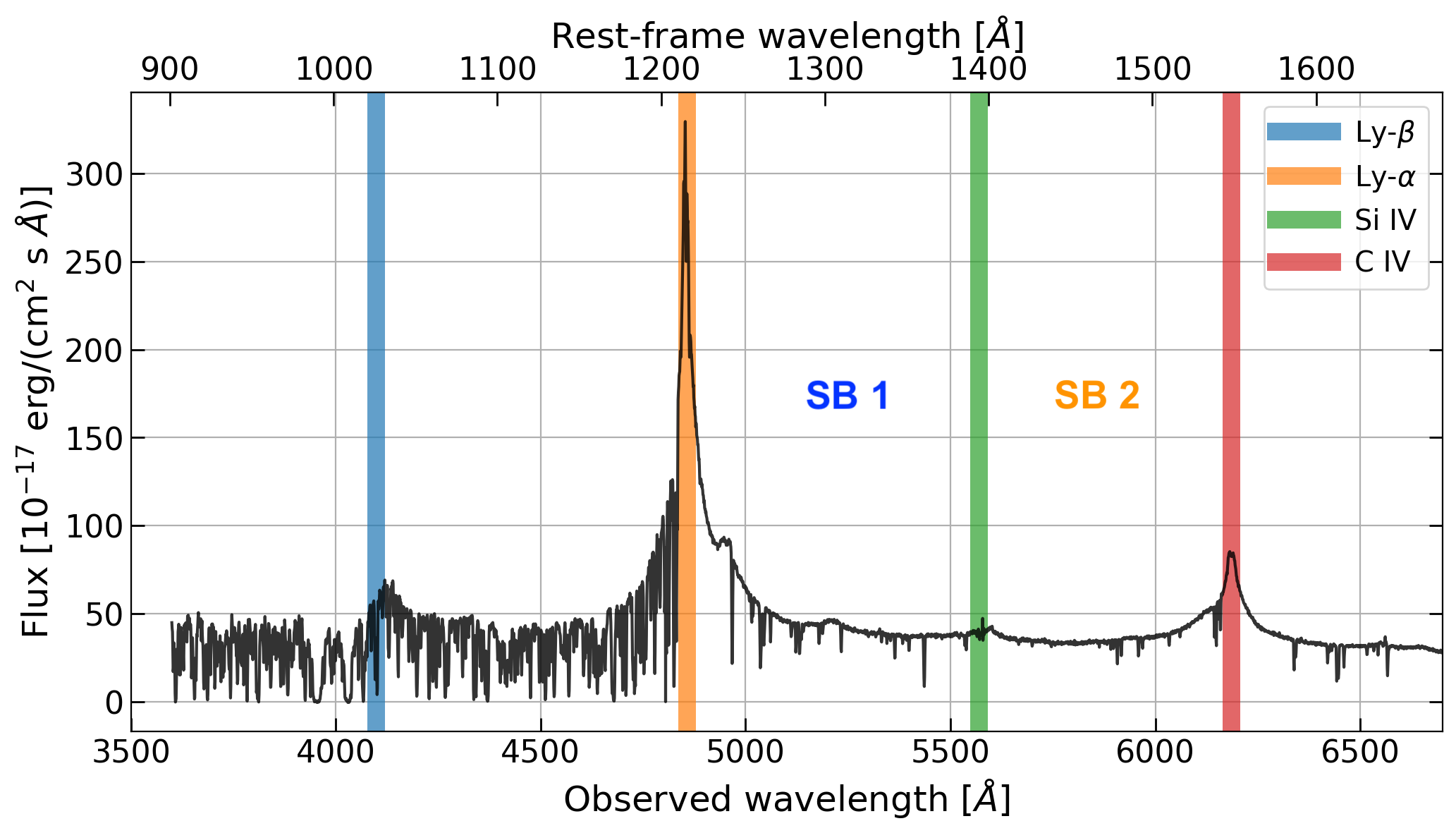}
    \caption{Quasar at $z=2.99$ observed during DESI survey validation (TargetID 39633362754732929).
    The Ly~$\alpha$ forest is typically studied between the Ly~$\alpha$ and Ly~$\beta$ emission lines.
    Absorption redward of Ly~$\alpha$ in the quasar rest-frame cannot be \ion{H}{I} from the IGM, but may be intervening metal systems such as \ion{Si}{IV} and \ion{C}{IV}.
    The regions from Ly~$\alpha$--\ion{Si}{IV} and from \ion{Si}{IV}--\ion{C}{IV} are called the "side bands" (SB) in 1D power spectrum studies, and are used to estimate the metal contamination in the forest.
    We call the Ly~$\alpha$--\ion{Si}{IV} region SB~1 and the \ion{Si}{IV}--\ion{C}{IV} region SB~2.
    }
    \label{fig:fuji_example_spectrum}
\end{figure*}

\begin{figure}
    \centering
    \includegraphics[width=\columnwidth]{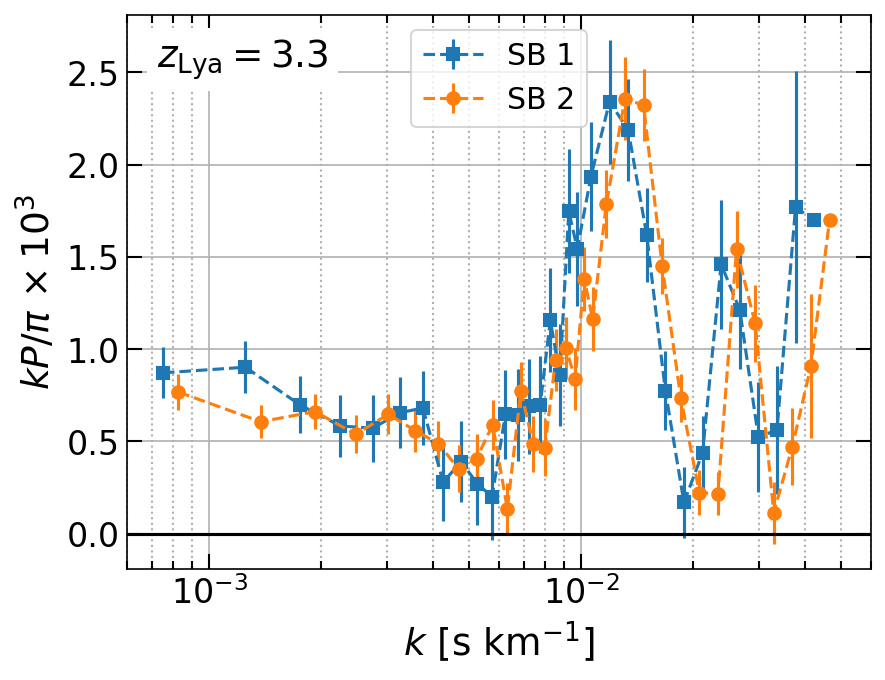}
    \caption{Measured side band power spectra from DESI early data at a Ly~$\alpha$ redshift of 3.3 (see Section~\ref{sec:data_analysis} for details).
    Large scales are contaminated by continuum errors, whereas small scales are limited by the spectrograph resolution.
    The \ion{C}{IV} doublet leaves a clear feature on the side band power spectrum as oscillations peaking at $k=0.013$\skm.}
    \label{fig:intro_sb_power_onez}
\end{figure}

The metals in the Universe are produced by star formation within galaxies and ejected into the circumgalactic medium (CGM) and the intergalactic medium (IGM) through various feedback processes \citep[see][for a review]{tumlinsonCircumgalacticMedium2017}.
The CGM is a multi-phase and complex medium with temperatures ranging from $10^4$~K to $10^6$~K \citep{andersonHotHalos2013}, whereas the IGM is typically associated with temperatures of $10^4$~K \citep{mcquinnEvolutionIntergalacticMedium2016}.
The rise and decline of strong \ion{Mg}{II} absorbers with redshift traces the cosmic star formation history, which suggests links between metals and the fuelling/feedback of star formation and galactic superwinds \citep{tumlinsonCircumgalacticMedium2017}.
Similarly, the increase in \ion{C}{IV} abundance in the IGM from $z\sim4.3$ to $z\sim2.4$ shows that some fraction of heavy elements seen at $z\sim 2.4$ must have been ejected from galaxies by that time \citep{simcoeCarbonContent2011}.
Different models that include galactic winds and Population III stars have different implications for metal enrichment and clustering in the inter- and circum-galactic media.

Metal line systems in the IGM and CGM have been observed by their absorption profiles in quasar spectra since 1980s, and have been studied extensively \citep{sargent_distribution_1980ApJS, sargentIVAbsorptionNew1988, steidel_high-redshift_1990, cowieMetallicityInternalStructure1995, ellisonEnrichmentHistoryIntergalactic2000, songailaMinimumUniversalMetal2001, pichon_clustering_2003, schayeMetallicityIntergalacticMedium2003, scannapiecoSourcesIntergalacticMetals2006, dodoricoRiseIvMass2010, cookseyPreciousMetalsSDSS2013, dodoricoMetalsIGMApproaching2013, boksenberg_sargent_properties_2015, hasanEvolutionIVAbsorbers2020}.
Typically, these systems are individually identified by visual inspection and/or by automated search routines.
Various quantities such as metallicity, column density or equivalent width distributions and cloud-cloud clustering are then measured from these individual detections.
These quantities tie metal systems to their local environment and are rich in information.
These studies usually focus on a small number of high-resolution, high signal-to-noise ratio (SNR) spectra because they require long observations with large telescopes. For example, \citet{hasanEvolutionIVAbsorbers2020} used 369 spectra to study the distribution and evolution of \ion{C}{IV} absorbers. 
Nevertheless, the field is not limited to small samples.
An exceptionally large study from \citet{cookseyPreciousMetalsSDSS2013} visually verified 16\,000 \ion{C}{IV} systems from approximately 100\,000 quasars in the Sloan Digital Sky Survey \citep[SDSS]{yorkSDSSTechinalSummary2000AJ} Data Release 7 \citep{abazajianTheSeventhDataReleaseSDSS2009ApJS, schneiderSDSSQuasarCatalogSeventh2010AJ}.

A key complication faced by line-identification studies is the identifier completeness (human or machine).
The completeness depends on SNR of the spectra and depth of the metal absorption, where confidence and completeness increase with both.
This bias is further complicated by the requirement that metal lines exist in an observed sight-line in the first place.
Long exposure times are typically focused on objects that manifest interesting features, resulting in a somewhat biased archive for SNR-limited samples.
In contrast, large spectroscopic surveys of quasars are largely free of bias with respect to intergalactic and circumgalactic medium properties.

Even though moderate resolution spectra from SDSS cannot measure individual weak column density systems, it can study these systems statistically.
Two current methods that allow such statistical studies of metals are the pixel optical depth (or pixel correlation search) method and stacked/composite spectra.
In the pixel optical depth method, the Ly~$\alpha$ optical depth in each pixel is compared to the optical depth at the corresponding metal-line wavelength \citep{cowieHeavyElementEnrichment1998, aguirreMetallicityPixelStatistics2002, schayeMetallicityIntergalacticMedium2003}.
The lack of homogeneity and cosmic variance challenges in small samples require modifications for pixel correlation searches to be suitable for SDSS spectra \citep{pieriSearchForOxygen2010}.
In the stacking method, the whole spectrum is shifted to the Ly~$\alpha$ absorber rest frame, stacked, and repeated for all Ly~$\alpha$ absorber detections for each quasar \citep{pieriCOMPOSITESPECTRUMSTRONG2010, pieriProbingCircumgalacticMedium2014, yangMetalLinesAssociated2022}.
The requirement for statistically significant absorber detections can further be relaxed to probe weak metal populations \citep{frankAgnosticStacking2018}.
These works derive the column densities of various metal species as a function of Ly~$\alpha$ absorber strength and redshift, which then can be used to study the chemical enrichment of the CGM and IGM.
These are promising methods that can provide metallicities for populations of different column densities while being statistically robust.

A complication for pixel optical depth and stacked/composite spectra is determining the thresholds for the Ly~$\alpha$ optical depth.
Such uncertainties at the detection limit make pixel studies difficult to interpret without mocks \citep{pieriPixelCorrelationOVI2004}.
These studies further require a significant signal (typically Ly~$\alpha$) with which to correlate.
\citet{frankAgnosticStacking2018} address the impact of stacking low probability noisy signal by attempting to measure the weak \ion{Ne}{VIII} signal with "agnostic stacking," 
in which all pixels that show apparent absorption regardless of their source transition are treated as \ion{Ne}{VIII}.
The resulting stacked signal is diluted by both pipeline noise and non-\ion{Ne}{VIII} absorbers, which then still requires an optimal selection for absorbers to maximize, though are capable of probing weak systems nonetheless.

\begin{table*}
    \centering
    \begin{tabular}{l|c|c|c|c|c|c}
        Ion & $\lambda_1$ [\AA] & $A_1$ [s$^{-1}$] & $f_1$ & $\lambda_2$ [\AA] & $A_2$ [s$^{-1}$] & $f_2$ \\
        \hline
        \ion{Si}{IV} & 1393.76 & 8.80$\times10^8$  & $0.513$ & 1402.77 & 8.63$\times10^8$ & $0.255$ \\
        \ion{C}{IV}  & 1548.20 & 2.65$\times10^8$ & $0.190$ & 1550.77 & 2.64$\times10^8$ & $0.0952$ \\
        \ion{Mg}{II} & 2795.53 & 2.60$\times10^8$ & $0.608$ & 2802.70 & 2.57$\times10^8$ & $0.303$ \\
        \ion{Fe}{II} & 2373.74 & 4.25$\times10^7$ & $0.0359$ & 2382.04 & 3.13$\times10^8$ & $0.320$ \\
        \ion{Fe}{II}-2 & 2585.88 & 8.94$\times10^7$ & $0.0717$ & 2599.40 & 2.35$\times10^8$ & $0.239$ \\
        \hline
    \end{tabular}
    \caption{Atomic data from  NIST \citep{NIST_ASD} for the ions in this study. The columns provide the doublet transition wavelengths $\lambda_*$, their spontaneous emission coefficients $A_*$, and their oscillator strengths $f_*$.}
    \label{tab:ion-transitions}
\end{table*}

We propose to mitigate these systematics by using 1D two-point statistics similar to large-scale structure studies while trading off the local environment information.
Our proposed method to study the metal abundance and clustering treats the flux as a continuous field and connects previous detection-based endeavours to the large-scale structure framework.
This method has the potential to measure the redshift evolution of cosmic metal abundance (e.g. $\Omega_{\ion{C}{IV}}(z)$) using large numbers of quasar spectra in a different way.
It does not require measurement of associated Ly~$\alpha$ absorption to measure the metal signal, which allows us to be sensitive to weaker absorption systems and to exploit lower SNR data.
Even though agnostic stacking allows a more direct inference of metal population properties,
our model is able to measure the metal signal mixed more fully within the noise and to measure metal clustering.
The future one-year and five-year data from Dark Energy Spectroscopic Instrument (DESI) \citep{leviDESIExperimentWhitepaper2013, desicollaborationDESIExperimentPart2016, abareshiOverviewInstrumentationDark2022} will provide hundreds of thousands of medium resolution spectra and will be an excellent data set to apply this framework.

The Ly~$\alpha$ forest is formed by neutral hydrogen in the inter- and circum-galactic media, and is observed only below the Ly~$\alpha$ transition line in the quasar rest frame.
Similarly, the \ion{C}{IV}, \ion{Si}{IV} and other ions create forests below their transition lines, which leave signals in the 1D power spectrum (\poned).
These regions are called "side bands" (SB) in Ly~$\alpha$ \poned studies, and are used to subtract the metal contamination from Ly~$\alpha$ forest measurements \citep{mcdonaldLyUpalphaForest2006, palanque-delabrouilleOnedimensionalLyalphaForest2013, chabanierOnedimensionalPowerSpectrum2019, karacayliOptimal1DLy2022}.
Figure~\ref{fig:fuji_example_spectrum} shows a quasar at $z=2.99$ from DESI early data.
The side bands are numbered starting at the Ly~$\alpha$ line, such that the Ly~$\alpha$--\ion{Si}{IV} region is called SB~1 and the \ion{Si}{IV}--\ion{C}{IV} region is called SB~2.
Much like there is \ion{Si}{IV} and \ion{C}{IV} absorption in Ly~$\alpha$ forest, there is \ion{Si}{IV} and \ion{C}{IV} absorption in SB~1, but no \ion{Si}{IV} absorption in SB~2.
Furthermore, the measured 1D flux field correlation function (power spectrum) in these side bands show clear peaks (oscillations) due to the doublet nature of the most dominant metal transitions.
As an illustration, Figure~\ref{fig:intro_sb_power_onez} clearly shows \ion{C}{IV} doublet oscillations that peak at $k=0.013$\skm in the side band \poned from DESI early data (see Section~\ref{sec:data_analysis} for details).

The flux correlation function was theoretically studied by \citet{hennawiProbingReionizationEarly2021} for \ion{Mg}{II} and \citet{tieConstrainingIGMEnrichment2022} for \ion{C}{IV} at high redshifts in the context of reionization with numerical simulations.
Here, we develop a theoretical model for the flux field that can be extended to any regime given certain statistical properties are known or can be calculated.
We use a simplified version of the absorber model of \citet{irsicAbsorberModelHalolike2018} which is analogous to the halo model of \citet{coorayHaloModelsLarge2002}.
We formulate the flux field two-point statistics in terms of the column density distribution, an effective Doppler parameter for temperature, and cloud-cloud clustering (analogous to halo-halo correlation function in halo model) of discrete systems.
We then apply our model to data.

This paper is organized as follows:
In section~\ref{sec:absorber_model} we introduce a toy model and then develop our full absorber model.
We detail our method and software used in Section~\ref{sec:method}. This includes a description of how we estimate the power spectrum and infer the model parameters. We describe the high-resolution spectra and DESI data in Section~\ref{sec:data_analysis}. 
We discuss our findings and some limitations of our model and data in Section~\ref{sec:discussion} and summarize our results in Section~\ref{sec:summary}.

\section{Absorber model}
\label{sec:absorber_model}
We start this section with a toy model for a single doublet profile, then move on to developing the full model.

\subsection{Toy model}
Let us build our intuition with a simple model by approximating the transition profile of a \emph{single} doublet with Gaussian functions $g(x)$ in the optically thin limit.
A doublet features two absorption lines, one of which is a factor of $r$ weaker.
For a doublet centred at $v$ with velocity separation $\mu$ between two lines, we can write the total profile $K$ and find its Fourier transform as follows:
\begin{align}
    K(v) & = g(v-\mu/2) + r g(v+\mu/2) \\
    \tilde{K}(k) & = \tilde{g}(k) \left(\mathrm{e}^{-\mathrm{i}k\mu/2} + r \mathrm{e}^{\mathrm{i}k\mu/2} \right)
\end{align}
The power spectrum will then be $P\propto \left| \tilde{K} \right|^2$.
\begin{align}
    P &= \left|\tilde{g}\right|^2 \left(1+r^2 +r \left(\mathrm{e}^{-\mathrm{i}k\mu} + \mathrm{e}^{\mathrm{i}k\mu} \right) \right)\\
    &= \left|\tilde{g}\right|^2 \left(1+r^2 + 2r\cos(k\mu) \right)
\end{align}
In this optically thin limit, the Doppler parameter $b$ is responsible for the line width, such that $\tilde{g}(k) = \tilde{g}(0)\mathrm{e}^{-k^2b^2/2}$.
Finally, $\tilde{g}(0) = \int \ddif v\, g(v) = \mathrm{EW}$ (equivalent width).
\begin{equation}
    P(k) \propto \mathrm{EW}^2 \left(1+r^2 + 2r\cos(k\mu) \right) \mathrm{e}^{-k^2b^2} \label{eq:single_doublet_power}
\end{equation}

This equation tells us that a single doublet feature produces oscillations in the power spectrum and the oscillation amplitude is proportional to the system's equivalent width two times the relative strength $r$ of the weaker transition.
Note that $r$ is purely determined by atomic physics when lines are unsaturated.
The Doppler parameter $b$ broadens the lines due to thermal motion and therefore suppresses the power on small scales.
Equation~(\ref{eq:single_doublet_power}) is a surprisingly good fitting function according to our preliminary tests; however, it is hard to interpret their results without building a sophisticated model.

Naturally, there could be multiple doublets in a given line of sight such that $K_\mathrm{tot}=\sum_i K_i$.
Then, the two-point function would be:
\begin{equation}
    \langle K_\mathrm{tot}K_\mathrm{tot}\rangle = \sum_i \langle K_i K_i \rangle + \sum_{i\neq j} \langle K_i K_j \rangle. \label{eq:toymodel_ktot}
\end{equation}
The second term vanishes when the doublets are uncorrelated, but they are in fact clustered as we discuss in the next section.

\subsection{Full model}
The halo model approach assumes that all particles reside in discrete dark matter halos. This model splits the statistics of the mass density field into two components. The small-scale statistics depends on the density distribution of individual halos (i.e., halo profile), and it is called the one-halo term. The large-scale statistics depends on the spatial distribution of halos (i.e., halo clustering), and this is called the two-halo term \citep{coorayHaloModelsLarge2002}.
Similar to the halo model, the absorber model decomposes the two-point flux statistics into contributions of absorption profiles from one or multiple ions, where the correlation function consists of one-absorber (1a) and two-absorber (2a) terms \citep{irsicAbsorberModelHalolike2018}.
The one-absorber term captures the doublet shape's correlation with itself (first term in equation~(\ref{eq:toymodel_ktot})), whereas the two-absorber term captures the clustering of two different systems (second term in equation~(\ref{eq:toymodel_ktot})).
The key difference of absorber model is that it does not build on actual dark matter halos, but uses the formalism of discrete tracers.

We start with expressions for the doublet absorption profile.
The optical depth of a transition line with wavelength $\lambda_{*}$, spontaneous emission coefficient $A_*$, oscillator strength $f_{*}$ and column density $N$ is given by the Voigt profile, which we approximate using the Tepper-García function $T(x; \sigma, \gamma)$ \citep{garciaVoigtProfileFitting2006a}:
\begin{equation}
    \tau_* = \frac{1}{\sqrt{\pi}} \left(\frac{\lambda_{*}}{b} \frac{\mathrm{km}}{\text{\AA}} \right) \left( \frac{N a_{*}}{10^{13}} \right)   T\left( \frac{\lambda}{\lambda_{*}}-1; \sigma_G, \gamma \right),
\end{equation}
where the Doppler parameter $b$ and the speed of light $c$ are in \kms, the wavelengths are in \AA\, and the column density $N$ is in cm$^{-2}$. 
The Gaussian spread is given by $\sigma_G=b/c$; 
the Lorentzian parameter is given by $\gamma = A_* \lambda_* \Gamma/ 4\pi c$, where $\Gamma \equiv 10^{-13}$~km~\AA$^{-1}$ is conversion coefficient between distance units; 
and $a_{*}=\pi q_e^2 f_{*}/m_e c=0.02654 f_{*}$~cm$^{2}$\,s$^{-1}$ in cgs units. 
Tepper-García function is given by the following analytic expression:
\begin{align}
    T(x; \sigma, \gamma) &= 
     \mathrm{e}^{-u^2} - \frac{a}{u^2\sqrt{\pi}} H\left(u\right)  \\
    H(u) &= \mathrm{e}^{-2u^2} \left( 4 u^4 + 7u^2 +4 + \frac{3}{2u^2} \right) - \frac{3}{2u^2} - 1,
\end{align}
where $u=x/\sigma$ and $a=\gamma/\sigma$.
We obtain ion transition wavelengths and their respective oscillator strengths from the National Institute of Standards and Technology (NIST)
Atomic Spectra Database\footnote{\url{https://physics.nist.gov/PhysRefData/ASD/lines_form.html}} \citep{NIST_ASD}.
Our values are written in Table~\ref{tab:ion-transitions}.
The normalized flux is given by $F=\mathrm{e}^{-\tau}$.
We also define $K \equiv 1-F$.

The absorption profile is the doublet shape such that the optical depths of the two transition lines are added: $\tau = \tau_1 + \tau_2$.
The final profile $K$ only depends on the Doppler parameter $b$ and the column density $N$ for a given ion:
\begin{equation}
    K(\lambda; b, N)=1-\exp\left[-\tau_1(\lambda; b, N) - \tau_2(\lambda; b, N) \right] \label{eq:doublet_K}.
\end{equation}
The rest are determined by atomic physics, which includes the relative strength $r$ (note $r$ also depends on $N$, i.e. saturation).

The spectra are reported in wavelength units, but we prefer to map the wavelength to velocity by $v=c\ln(\lambda/\lambda_{\mathrm{pivot}})$.
This mapping originates from \poned measurements of logarithmically spaced SDSS spectra,
but note that it does not correspond to a physical velocity in space. 
The exceptional feature of these velocity units is that the doublet separations do not depend on the pivot or absorber redshifts.
For example, all \ion{C}{IV} doublets at all redshifts occur at a separation of $\Delta v \approx 500$\kms.

The flux correlation function is defined as $\xi(v)\equiv\langle\delta_F(v')\delta_F(v'+v)\rangle$, where $\delta_F\equiv F/\overline{F} - 1$.
However, the metals in the side band are mostly weak and sparse, so that we assume $\overline{F}=1$ in the side bands.
The mean flux can be calculated and added to this model, but the deviations from one are small, such that mean flux errors can be marginalized out like large-scale continuum errors.
Therefore, our flux correlation function definition is equivalent to $\xi(v)=\langle K(v') K(v'+v)\rangle$.
Then, the absorber model correlation function is given by integrations over the column density distribution $f(N)$.
\begin{align}
    \xi_{1a}(v) &= \int \ddif N_i f(N_i) \int \ddif v' K_i(v') K_i(v'+v) \label{eq:xi1a}\\
    \xi_{2a}(v) &= \int \ddif N_i \ddif  N_j  f(N_i) f(N_j) \nonumber \\
        &\quad \times \int \ddif x \ddif v' K_i(v') K_j(v'+x+v) \xi_{cc} (x; N_i, N_j), \label{eq:xi2a}
\end{align}
where $K_i = K_i(v; b_\mathrm{eff}, N_i)$ as given in equation~(\ref{eq:doublet_K}) and $\xi_{cc}$ is the two-point cross-correlation function between systems of column density $N_i$ and $N_j$. 
As we stated previously, the one-absorber term $\xi_{1a}$ and two-absorber term $\xi_{2a}$ correspond to the first and second terms in equation~(\ref{eq:toymodel_ktot}) respectively.
For simplicity, we assume a single effective Doppler parameter $b_\mathrm{eff}$.
However, it is easy to extend this model for some $b$ distribution by replacing $\int \ddif N f(N) \rightarrow \int \ddif N \ddif b  f(N, b)$.
Since the physically additive quantity is the optical depth instead of $K$, there are higher order contributions to the two-point function \citep{irsicAbsorberModelHalolike2018}.
We ignore those terms here.
Note that the velocity integrations can be implemented by Fast Fourier Transforms (FFT) on a fine, equally spaced grid.

We perform an empirical study based on previous line identification studies for column density distribution $f(N)$ and cloud-cloud clustering $\xi_{cc}$, but our long term goal is to independently constrain all parameters of the model.
As a forewarning to Section~\ref{sec:method}, we will keep $\xi_{cc}$ fixed and limit free parameters to an effective Doppler parameter $b_\mathrm{eff}$ and an amplitude scaling of $f(N)$ for each ion.

\begin{table}
    \centering
    \begin{tabular}{l|c|c|c|c|c}
        Ion & $f_0$ & $\alpha$ & $\xi_0$ & $r_0$ [\kms] & $\gamma$\\
        \hline
        \ion{C}{IV} & -12.7 & 1.8 & 57 $\pm$ 8 & 90 $\pm$ 16 & 1.8 $\pm$ 0.2 \\
        \ion{Si}{IV} & -13.5 & 1.7 & 130 $\pm$ 55 & 45 $\pm$ 22 & 1.6 $\pm$ 0.2 \\
        \ion{Mg}{II} & -13.2 & 1.6 & 170 $\pm$ 20 & 125 $\pm$ 15 & 2.6 $\pm$ 0.2 \\
        \ion{Fe}{II} & -13.4 & 1.7 & 270 $\pm$ 55 & 100 $\pm$ 15 & 3.1 $\pm$ 0.3 \\
        \hline
    \end{tabular}
    \caption{Fiducial parameter values for the column density distribution and small-scale clustering.
    These values rely on \citet{scannapiecoSourcesIntergalacticMetals2006}, where \ion{C}{IV} and \ion{Si}{IV} systems are in the $1.5<z<3.1$ redshift range, 
    whereas \ion{Mg}{II} and \ion{Fe}{II} systems are in the $0.4<z<1.9$ redshift range.
    The $f_0$ and $\alpha$ values are the best fits by \citeauthor{scannapiecoSourcesIntergalacticMetals2006} for our fiducial column density distribution and 
    $\xi_0$, $r_0$ and $\gamma$ are our  best-fitting values to equation~(\ref{eq:xicc_fn}) using measurements in \citeauthor{scannapiecoSourcesIntergalacticMetals2006}.
    The mean redshifts of \ion{C}{IV} and \ion{Si}{IV} are $z\approx 2.3$, and are $z=1.15$ for \ion{Mg}{II} and \ion{Fe}{II}.
    We ignore the redshift evolution of this function for simplicity. 
    }
    \label{tab:fiducial_fN_xicc}
\end{table}

\begin{figure}
    \centering
    \includegraphics[width=\columnwidth]{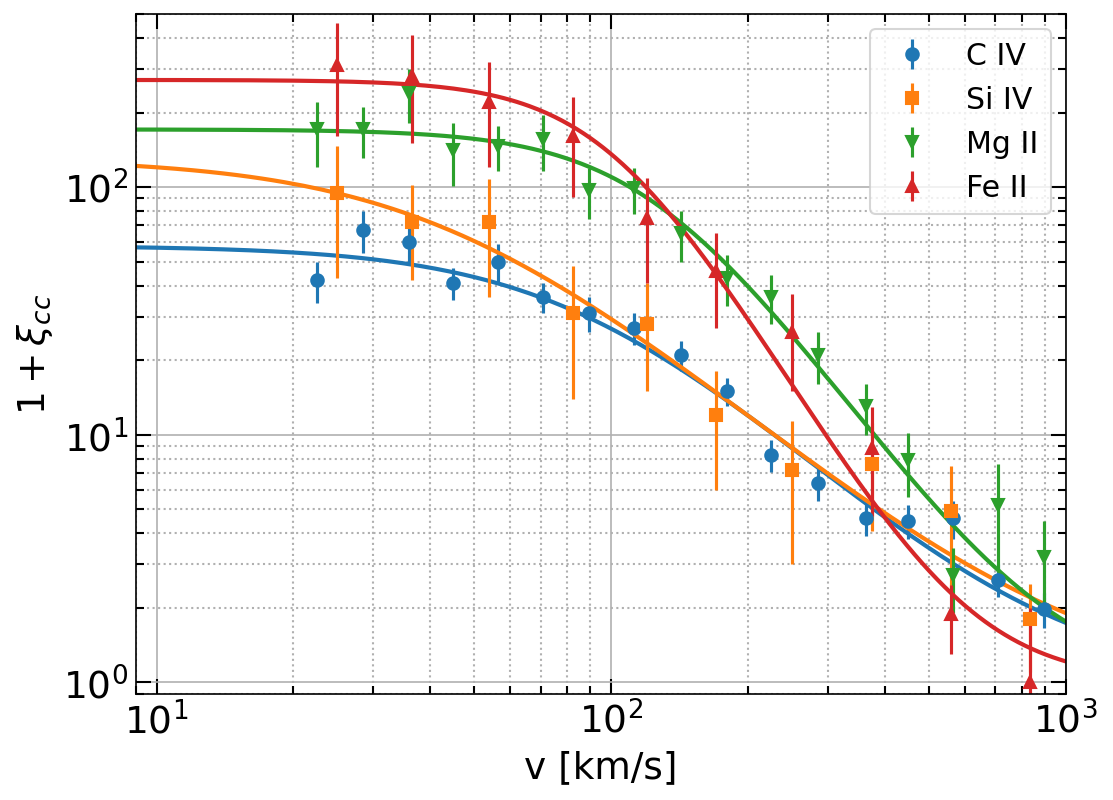}
    \caption{Small-scale clustering of ions ({\it filled points}) measured by \citet{scannapiecoSourcesIntergalacticMetals2006}.
    The solid lines are calculated with equation~(\ref{eq:xicc_fn}) and the  best-fitting parameter values listed in Table~\ref{tab:fiducial_fN_xicc} for each ion.
    }
    \label{fig:xicc_bestfit}
\end{figure}

For column density distribution $f(N)$, we will use the  best-fitting parameters from \citet{scannapiecoSourcesIntergalacticMetals2006} for the functional form:
\begin{equation}
    f(N) = 10^{f_0} \left(\frac{N}{10^{13} \mathrm{cm}^{-2}} \right)^{-\alpha}
\end{equation}
The numerical values of $f_0$ and $\alpha$ are in Table~\ref{tab:fiducial_fN_xicc}.
In our analysis, we are keeping $\alpha$ fixed, and fit for deviations from $f_0$ for each ion.
Since both the 1a and 2a terms directly scale with the amplitude of $f(N)$, this makes the fitting simple.
We first calculate templates for the fiducial values, then scale each template by $10^{A_{\mathrm{ion}}}$, where $A_{\mathrm{ion}}$ is the fitting parameter (not to be confused with the spontaneous emission coefficient $A_*$).
Therefore, $A_\mathrm{ion}=0$ means agreement with \citeauthor{scannapiecoSourcesIntergalacticMetals2006}.
Furthermore, the column density distribution is usually given per column density per redshift path $X$ \citep{songailaMinimumUniversalMetal2001, scannapiecoSourcesIntergalacticMetals2006}. 
We need the column density distribution per velocity for our model, so we apply the following transformation to the measurements in literature.
\begin{equation}
    f_v(N) = f_X(N) \fullderiv{X}{z} \fullderiv{z}{v} = f_X(N) \frac{(1+z)^2}{E(z)} \frac{1+z}{c} \label{eq:fN_z}
\end{equation}
where $E^2(z)\equiv \Omega_\Lambda + \Omega_m (1+z)^3$. We assume a flat $\Lambda$CDM cosmology with $\Omega_m=0.315$ \citep{collaborationPlanck2018Results2020}.

\begin{figure}
    \centering
    \includegraphics[width=\columnwidth]{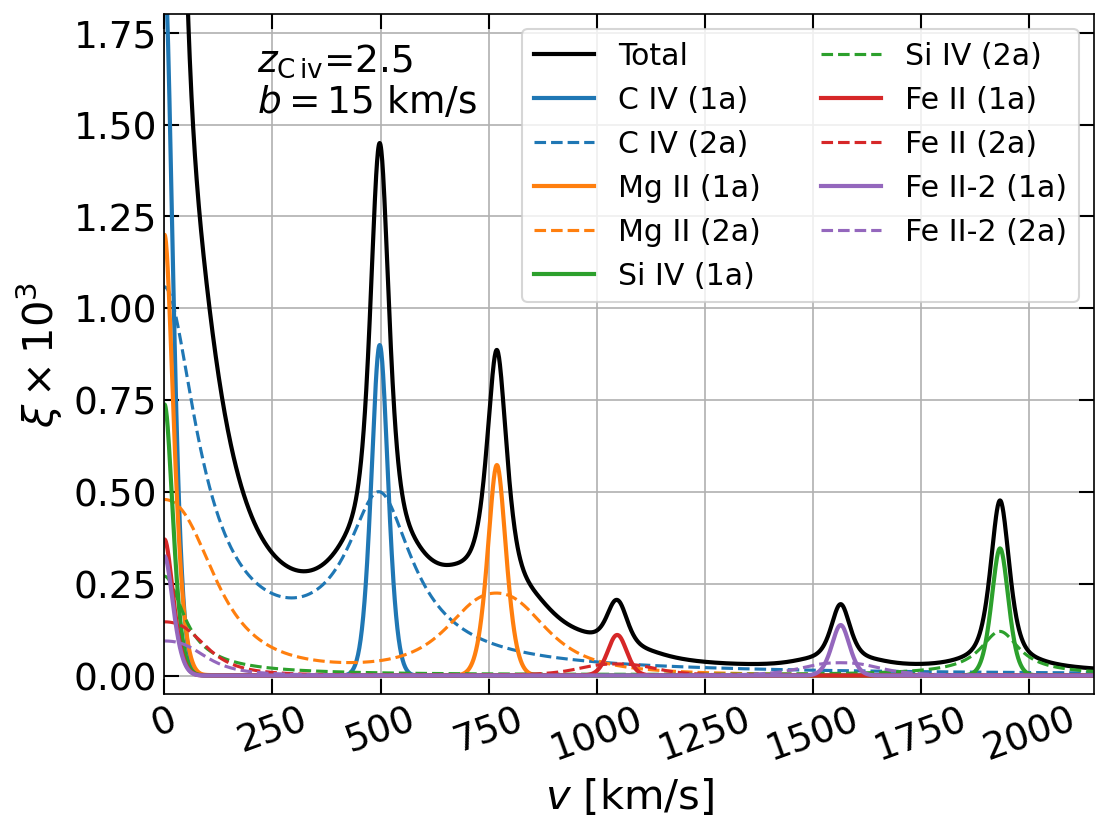} \\
    \includegraphics[width=\columnwidth]{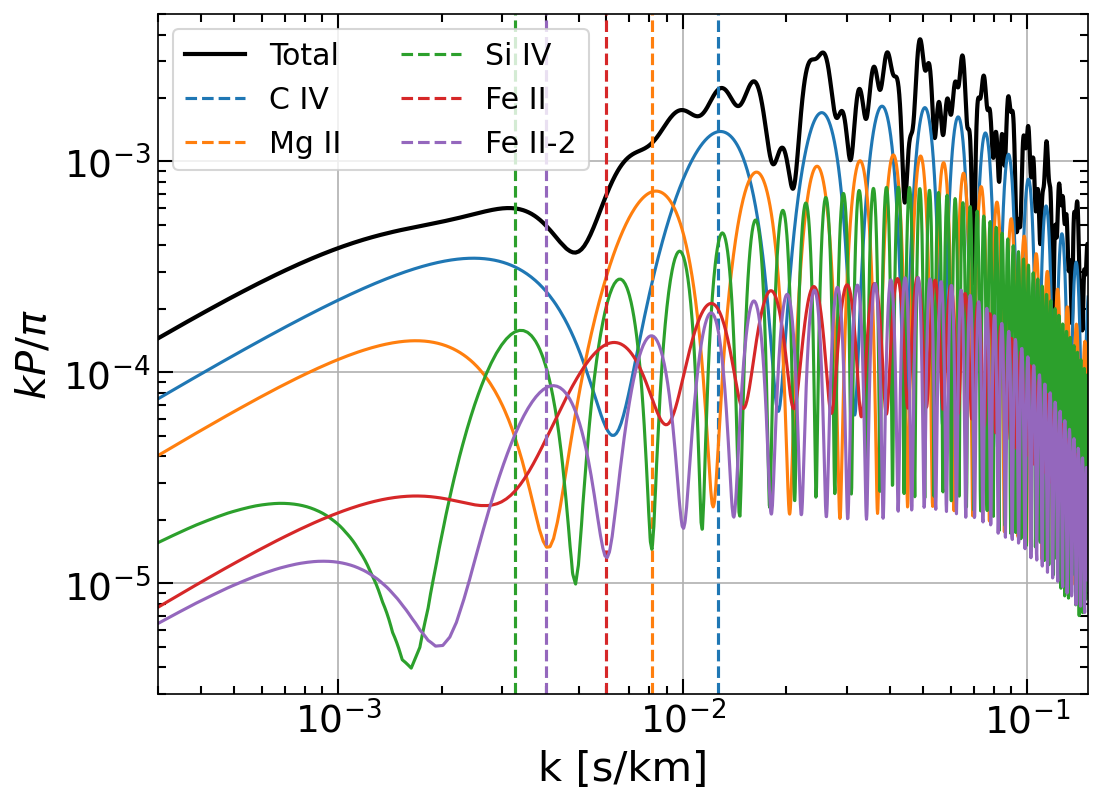}
    \caption{Two-point functions calculated from purely fiducial values for $b_\mathrm{eff}=15$\kms and \ion{C}{IV} redshift $z_\mathrm{\ion{C}{IV}}=2.5$.
    Other ions are at different redshifts according to their transition wavelengths: $z_\mathrm{\ion{Si}{IV}}=2.9$, 
    $z_\mathrm{\ion{Fe}{II}}=1.3$, 
    $z_\mathrm{\ion{Fe}{II}-2}=1.1$,
    $z_\mathrm{\ion{Mg}{II}}=0.9$. 
    (Top) Correlation function. Note the 2a term broadens the peaks and contributes approximately 50\% of the signal.
    (Bottom) Power spectrum. Localized peaks in the correlation function become modulated in the power spectrum with many overlapping oscillations.
    Even though some features remain, the maxima points (dashed lines, defined as $2\pi/\mu$) get significantly blurred.
    These features will be further blended when the power spectrum is measured in bins that are averages over certain $k$ ranges.
    }
    \label{fig:model_fid}
\end{figure}

\begin{figure}
    \centering
    \includegraphics[width=\columnwidth]{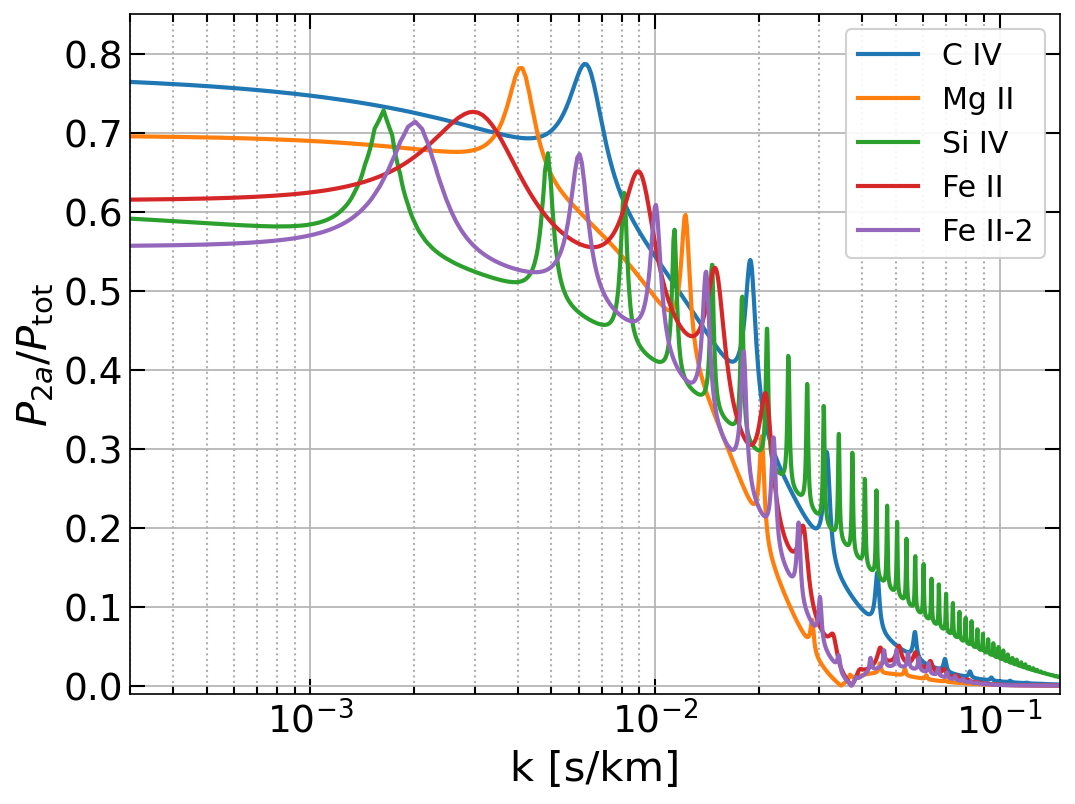}
    \caption{Relative contributions of the 2a terms to the power spectrum for each ion individually.
    The pivot \ion{C}{IV} redshift is $z_\mathrm{\ion{C}{IV}}=2.5$ as in Fig.~\ref{fig:model_fid}.
    Surprisingly, most of the signal comes from the 2a term for all ions for scales $k \lesssim 0.01$\skm.
    However, these results heavily rely on $\xi_{cc}$, which carry uncertainties as large as 40\%.}
    \label{fig:ratio2aterm}
\end{figure}

For our fiducial small-scale cloud-cloud clustering, we again use the measurements of \citet{scannapiecoSourcesIntergalacticMetals2006}, and fit the following function for each ion:
\begin{equation}
    \xi_{cc}(v) = \frac{\xi_0}{1+(v/r_0)^\gamma}\label{eq:xicc_fn},
\end{equation}
where $\xi_0$ is the clustering amplitude, $r_0$ is the correlation length and $\gamma$ is the slope.
All three are fitting parameters.
Figure~\ref{fig:xicc_bestfit} shows the \citeauthor{scannapiecoSourcesIntergalacticMetals2006} measurements and our best fit.

\citeauthor{scannapiecoSourcesIntergalacticMetals2006} find some evidence for $\xi_\mathrm{\ion{C}{IV}}$ redshift evolution, but
for this preliminary analysis we ignore the redshift evolution of this clustering.
\citet{dodoricoRiseIvMass2010} further indicate that $\xi_{cc}$ is in fact $N$ dependent, where higher column density systems are more clustered, but we again ignore the $N$ dependence for simplicity.
The best-fitting parameters are listed in Table~\ref{tab:fiducial_fN_xicc}.
Figure~\ref{fig:xicc_bestfit} shows the best-fitting curve as solid lines.

\subsection{Illustration of the model}
An illustration of the final model is shown in Figure~\ref{fig:model_fid}. 
We use the purely fiducial values outlined above, and pick a 
pivot \ion{C}{IV} redshift of $z_\mathrm{\ion{C}{IV}}=2.5$.
Other ions are at different redshifts according to their transition wavelengths: $z_\mathrm{\ion{Si}{IV}}=2.9$, 
$z_\mathrm{\ion{Fe}{II}}=1.3$, 
$z_\mathrm{\ion{Fe}{II}-2}=1.1$,
$z_\mathrm{\ion{Mg}{II}}=0.9$. 
They all follow the correct scaling with respect to equation~(\ref{eq:fN_z}).
The top panel shows the correlation function for $b=15$\kms, which would be expected from purely thermal broadening.
The 2a term broadens the peaks as expected and contributes approximately 50\% of the signal.
The corresponding power spectrum is plotted in the bottom panel.
Here the 1a and 2a terms are summed for each ion.
Localized peaks manifest as oscillations in the power spectrum, similar to baryon acoustic oscillations.
Since the final power is the sum of multiple sources,
the total power is blurred.
These features will be further blended as we measure the power spectrum in bins that are averages over certain $k$ ranges.
For visual guidance, we mark maxima points ($2\pi/\mu$) as dashed lines.

Figure~\ref{fig:ratio2aterm} shows the relative contribution of the 2a terms to the total power spectrum at the same
redshifts as before.
We find that most of the signal comes from the 2a term for all ions on scales of $k \lesssim 0.01$\skm.
These results heavily rely on $\xi_{cc}$, which carries uncertainties as large as 40\%. 
We do not consider these errors here, and hence the dominance of the 2a term is most convincing for \ion{C}{IV} and \ion{Mg}{II}.
One could introduce a separate fitting parameter for the $\xi_{cc}$ amplitude $10^{D_{\mathrm{ion}}}$.
Such a fitting function for an ion would be
\begin{equation}
    \xi = 10^{A_{\mathrm{ion}}} \xi_{1a} + 10^{2A_{\mathrm{ion}}+D_{\mathrm{ion}}} \xi_{2a} \label{eq:fitting_xiall},
\end{equation}
where $D_{\mathrm{ion}}$ is responsible for quantifying the uncertainties in the amplitude of $\xi_{cc}$ as well as its redshift evolution; however, this additional parameter could be degenerate with $A_{\mathrm{ion}}$.

\begin{figure}
    \centering
    \includegraphics[width=\columnwidth]{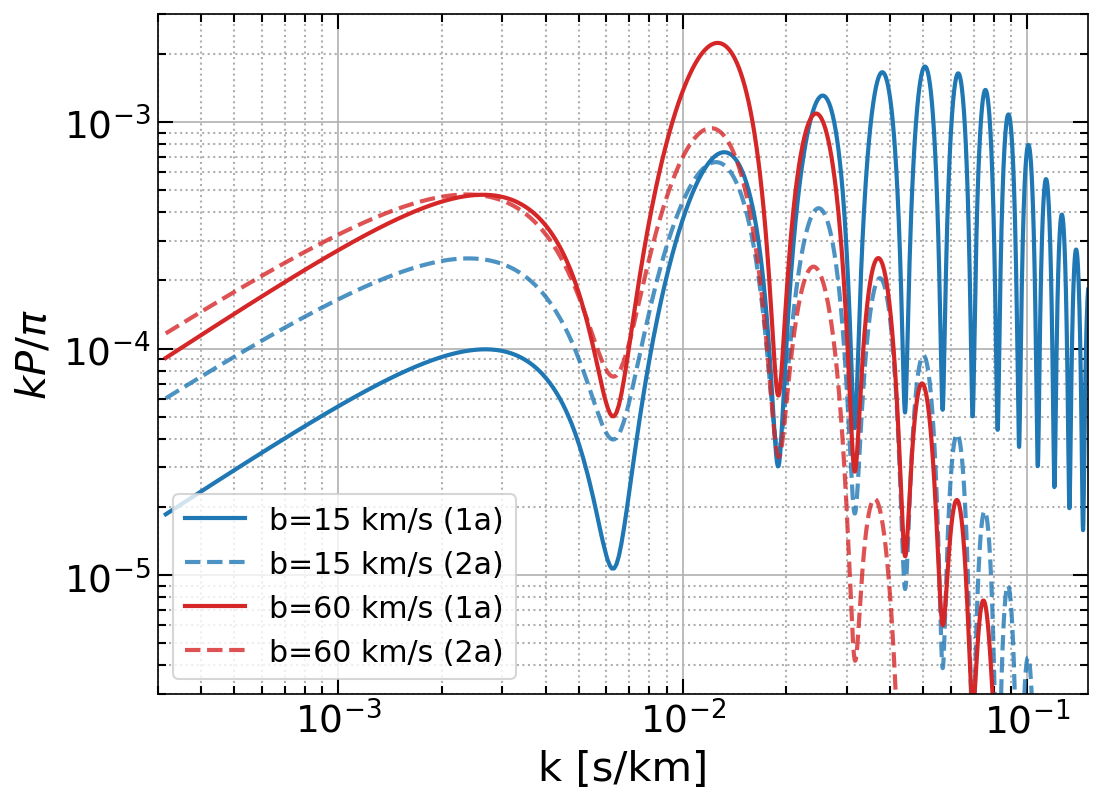}
    \caption{Comparison of two different Doppler $b$ values for \ion{C}{IV} at $z=2.5$.
    The solid lines represent the 1a term, whereas the dashed lines represent the 2a term.
    The lower $b=15$\kms yields a lower power spectrum (blue) at large scales than the higher $b=60$\kms value; 
    however, higher $b$ has less power at small scales as expected.
    }
    \label{fig:comp-b-civ-theory}
\end{figure}

Figure~\ref{fig:comp-b-civ-theory} shows the power spectrum for low and high $b$ values for \ion{C}{IV} at $z=2.5$.
Higher $b$ values bring the suppression to smaller $k$ values, but
also increase the amplitude at large scales.
The result is that the impact of the Doppler parameter is not trivial even for the 1a term.

\begin{figure}
    \centering
    \includegraphics[width=\columnwidth]{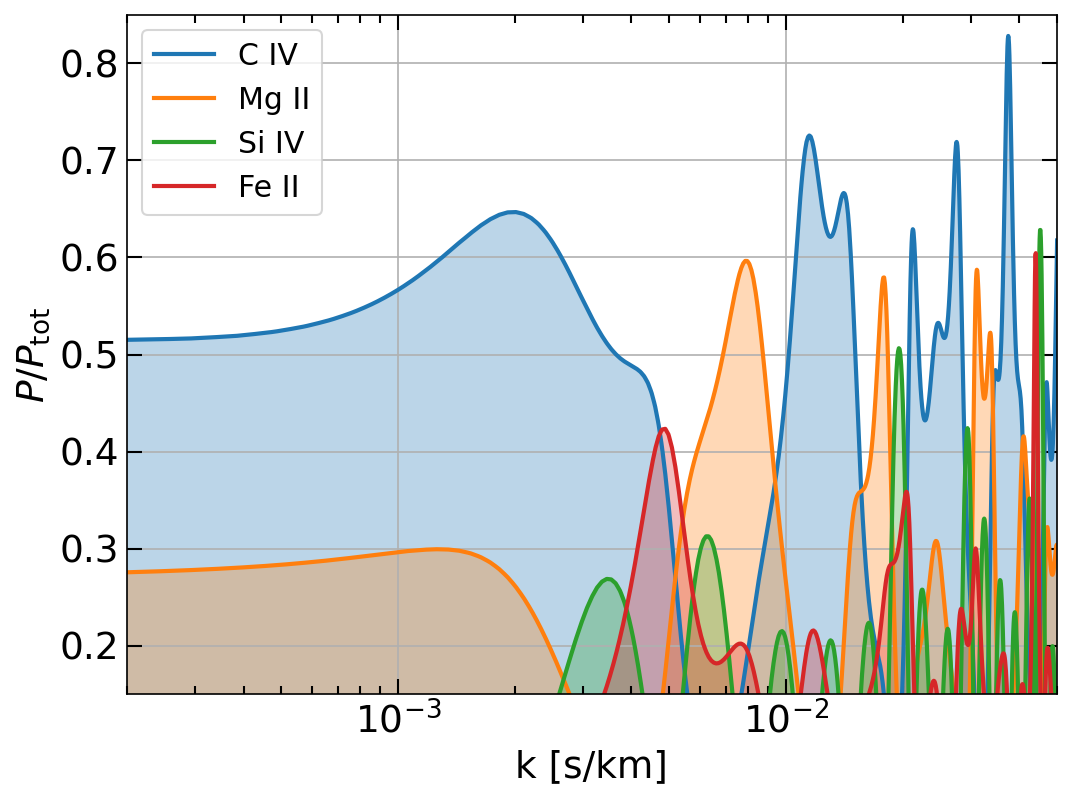}
    \caption{Contribution of each ion to the total power spectrum, which is the sum of the 
    1a and 2a terms.
    The majority of the signal comes from \ion{C}{IV} and \ion{Mg}{II} on most scales.
    }
    \label{fig:power_allions_ratio}
\end{figure}

Finally, Figure~\ref{fig:power_allions_ratio} shows the relative contribution of each ion to the power spectrum.
The 1a and 2a terms are summed up as in the bottom panel of Figure~\ref{fig:model_fid} and the majority of the signal comes from \ion{C}{IV} and \ion{Mg}{II} on most scales.
Other ions also exceed 20\% of the total power threshold at their distinctive scales.

\begin{figure}
    \centering
    \includegraphics[width=\columnwidth]{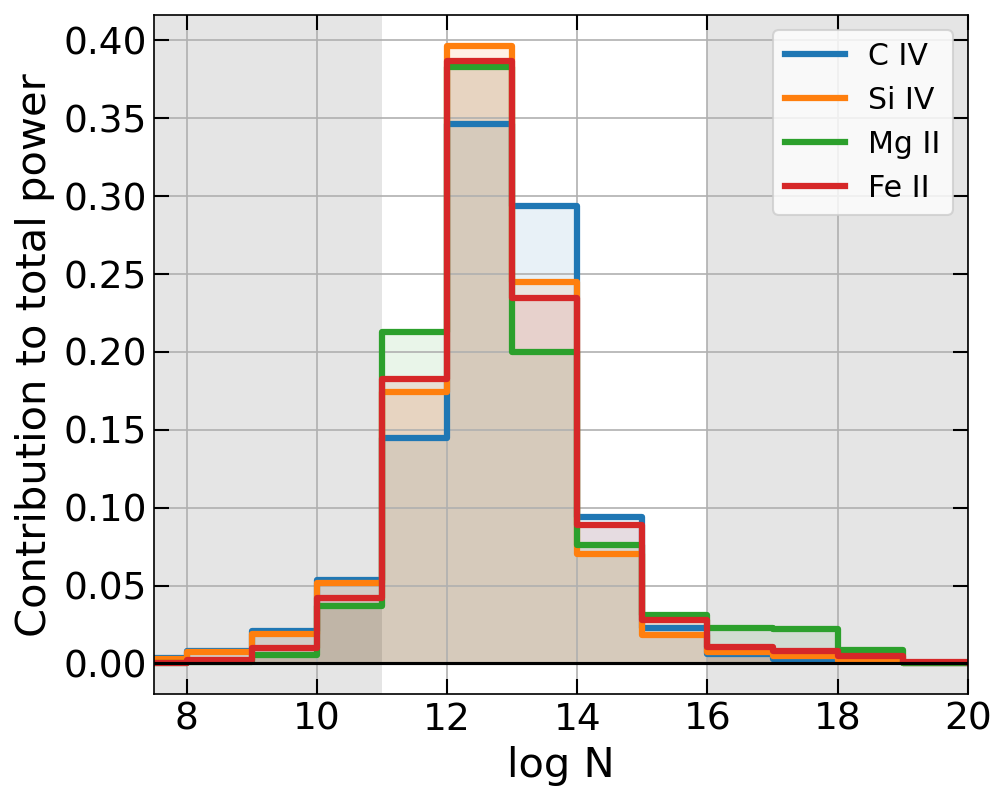}
    \caption{Contributions from each $\log N$ bin to the total power.
    We integrate each power spectrum from $k=0.001$\skm to $k=0.1$\skm and show the ratios.
    Approximately more than 90\% of the power comes from our nominal range 11--16 (unshaded region).
    }
    \label{fig:n_dependence}
\end{figure}

\subsection{Dependence on column density cuts}
\label{subsec:dependence_on_n}
We calculate the integrals over the column density distributions in equations~(\ref{eq:xi1a}) and (\ref{eq:xi2a}) within the measurement range of \citet{scannapiecoSourcesIntergalacticMetals2006} for $f(N)$ and $\xi_{cc}$, which is between $10^{11}$ and $10^{16}$.
However, as \citet{cookseyPreciousMetalsSDSS2013} note, a simple power law for $f(N)$ is divergent for quantities such as $\Omega_{\mathrm{\ion{C}{IV}}}$.
We address this in Section~\ref{sec:discussion}.
Fortunately, divergence is not an issue for our model.
Figure~\ref{fig:n_dependence} shows how much each logarithmic column density bin size of 0.1 contributes to the total power for each ion.
We integrate each power spectrum from $k=0.001$\skm to $k=0.1$\skm for the same 
pivot \ion{C}{IV} redshift $z_\mathrm{\ion{C}{IV}}=2.5$ and $b=15$\kms and show the ratios.
The excluded regions (shaded gray) contributes $\lesssim 10\%$ to the total signal.

We can intuitively understand this dependence considering only the one-absorber term.
In the optically thin limit where $\tau\propto N$ and $\langle KK \rangle \propto N^2$, the contribution of each column density to the two-point statistics from equation~(\ref{eq:xi1a}) reads as:
\begin{equation}
    \frac{\Delta \xi_{1a}}{\Delta \log N} \propto N^{3-\alpha}.
\end{equation}
According to this equation, the power contribution per $\log N$ climbs with column density $N$ for $\alpha<3$ and declines for $\alpha>3$.
All metals we consider in this work have $\alpha\sim 1.7$, so their contribution to the power spectrum increases with column density.
This contribution will eventually turn over as saturation fixes the power in $\langle KK \rangle=\mathrm{const}$ and the power contribution becomes $\Delta \xi_{1a}/\Delta \log N \propto N^{1-\alpha}$ for $\alpha>1$.
Note that the integration does not converge at high $N$ for $\alpha<1$.

Individual detection studies \citep[e.g.,][]{cookseyLastEightBillion2010, hasanEvolutionIVAbsorbers2020} on high-resolution spectra report only 50\% completeness at $\log N > 13$, whereas Figure~\ref{fig:n_dependence} shows that $\log N \approx 12-13$ systems primarily contribute to the total power in our model. High-resolution data sets \citep[e.g.,][]{omearaSecondDataRelease2017, murphyUVESSpectralQuasar2019} have SNR larger than five per pixel of 2.5\kms for most of their spectra calculated at 1450~\AA\ in the quasar’s rest frame. They also possess incredibly high-quality quasar spectra with SNR over 100 per pixel. In contrast, DESI is a medium resolution instrument with median SNR of two per pixel of 0.8~\AA, which is approximately 40\kms at 1450~\AA\ in the rest-frame for a quasar at $z=3$. Scaling this median SNR ratio to 2.5\kms pixel size gives an SNR of 0.5, which is significantly smaller than the mean SNR of the high-resolution data sets. Yet this difference is offset by the substantially larger number of DESI spectra. The early DESI data we present in Section~\ref{subsec:desi_edr} has over 40\,000 quasars, and the final sample will have over a million at sufficiently high redshift for this work. Even though the individual DESI spectra are well below the SNR threshold for individual detection limit of low column density systems, our model relies on the collective signal of these smaller systems in the power spectrum.

\section{Method}
\label{sec:method}
\subsection{Power spectrum estimation}
We measure \poned using the quadratic maximum likelihood estimator (QMLE).
QMLE works in real space (instead of Fourier space) to estimate the power spectrum, and therefore is not biased by gaps in the spectra, allows weighting by the pipeline noise, and accounts for the intrinsic Ly~$\alpha$ large-scale structure correlations.
We refer the reader to \citet{karacayliOptimal1DLy2020} and \citet{karacayliOptimal1DLy2022} for our development process and application to high-resolution spectra.
DESI related updates are detailed in a companion paper that measures the 1D Ly~$\alpha$ forest flux power spectrum from early DESI observations (Kara{\c c}ayl{\i} et al., in preparation).
We present a short summary of the relevant features in this section.

An important feature of our QMLE implementation is estimating deviations from a fiducial power spectrum such that $P(k, z) = P_{\mathrm{fid}}(k, z) + \sum_{m,n} w_{(mn)}(k, z) \theta_{(mn)}$, where we adopt top-hat $k$ bands with $k_n$ as bin edges and linear interpolation for $z$ bins with $z_m$ as bin centres.
We use the following fitting function to characterise the fiducial power spectrum:
\begin{equation}
    \label{eq:pd13_fitting_fn}\frac{kP(k, z)}{\pi} = A \frac{(k/k_0)^{3 +n + \alpha\ln k/k_0}}{1+(k/k_1)^2} \left(\frac{1+z}{1+z_{0}}\right)^{B + \beta\ln k/k_0},
\end{equation}
where $k_{0} = 0.009$\skm and $z_{0}=3.0$ \citep{palanque-delabrouilleOnedimensionalLyalphaForest2013, karacayliOptimal1DLy2020, karacayliOptimal1DLy2022}.
This fitting function is sufficient for a baseline estimate as $P_{\mathrm{fid}}$, which in turn can be used to weight pixels, but does not capture all scientific information in \poned.

Given a collection of pixels representing normalized flux fluctuations $\bm{\delta}_F$, the quadratic estimator is formulated as follows:
\begin{equation}
    \label{eq:theta_it_est}\hat \theta^{(X+1)}_{\alpha} = \sum_{\alpha'} \frac{1}{2} F^{-1}_{\alpha\alpha'}(d_{\alpha'} - b_{\alpha'} - t_{\alpha'}),
\end{equation}
where $X$ is the iteration number and 
\begin{align}
    \label{eq:data_dn} d_{\alpha} &= \bm{\delta}_F^\mathrm{T} \mathbf{C}^{-1}\mathbf{Q}_{{\alpha}} \mathbf{C}^{-1} \bm{\delta}_F, \\
    \label{eq:noise_bn}b_{\alpha} &= \Tr(\mathbf{C}^{-1}\mathbf{Q}_{\alpha} \mathbf{C}^{-1}\mathbf{N}), \\
    \label{eq:signalfid_tn}t_{\alpha} &= \Tr(\mathbf{C}^{-1}\mathbf{Q}_{\alpha} \mathbf{C}^{-1}\mathbf{S}_{\mathrm{fid}}),
\end{align}
where the covariance matrix $\mathbf C \equiv \langle\bm{\delta}_F\bm{\delta}_F^T\rangle$ is the sum of the signal and noise as usual, $\mathbf C = \mathbf{S}_{\mathrm{fid}} + \sum_{\alpha} \mathbf{Q}_{\alpha} \theta_{\alpha}+\mathbf{N}$, $\mathbf{Q}_{\alpha} = \partial \mathbf{C} / \partial \theta_{\alpha}$ and the estimated Fisher matrix is
\begin{equation}
    \label{eq:fisher_matrix}F_{\alpha\alpha'} = \frac{1}{2} \Tr(\mathbf{C}^{-1}\mathbf{Q}_{\alpha} \mathbf{C}^{-1} \mathbf{Q}_{\alpha'} ).
\end{equation}
The covariance matrices on the right hand side of equation~(\ref{eq:theta_it_est}) are computed using parameters from the previous iteration $\theta_{\alpha}^{(X)}$. Assuming different quasar spectra are uncorrelated, the Fisher matrix $F_{\alpha\alpha'}$ and the expression in parentheses in equation~(\ref{eq:theta_it_est}) can be computed for each quasar, then accumulated, i.e. $\mathbf{F}=\sum_q\mathbf{F}_{q}$ etc. 

DESI spectral extraction is built on an improved spectro-perfectionism algorithm \citep{boltonSpectroPerfectionismAlgorithmicFramework2010, guySpectroscopicDataProcessingPipeline2022}.
Spectro-perfectionism produces a resolution matrix $\mathbf{R}$ associated with each spectrum that is based on the spectrograph resolution as well as the noise properties of each spectrum, and captures the wavelength-dependent resolution on the same discrete wavelength bins as the spectrum.
The observed signal becomes a matrix-vector multiplication.
\begin{equation}
    \bm{\delta}_R = \mathbf{R} \bm{\delta}
\end{equation}
Our quadratic estimator naturally incorporates this matrix and deconvolves it from power spectrum measurements.
The signal $\mathbf{S}$ and derivative matrices $\mathbf{Q}$ are given by the following expressions:
\begin{align}
    \mathbf{S}_R &= \langle \bm{\delta}_R \bm{\delta}_R^T \rangle 
    = \mathbf{R} \mathbf{S} \mathbf{R}^T \\
    \mathbf{Q}^\alpha_R &= \mathbf{R} \mathbf{Q}^\alpha \mathbf{R}^T,
\end{align}
where the subscript $R$ denotes smoothed matrices, and matrices without a subscript are given by:
\begin{equation}
    S_{ij}^{\mathrm{fid}} = \int_0^\infty \frac{dk}{\pi} \cos(k v_{ij}) P_{\mathrm{fid}}(k, z_{ij}),
\end{equation}
where $v_{ij}\equiv v_i - v_j$ and $1 + z_{ij} \equiv \sqrt{(1+z_i)(1+z_j)}$, and the derivative matrix for redshift bin $m$ and wavenumber bin $n$ is
\begin{equation}
    Q_{ij}^{(mn)} = I_m(z_{ij}) \int_{k_n}^{k_{n+1}} \frac{dk}{\pi} \cos(kv_{ij}),
\end{equation}
where $I_m(z)$ is the interpolation kernel, which is one when $z=z_m$ and zero when $z=z_{m\pm1}$. 
We compute these matrices for as many redshift bins as necessary for a given spectrum.

\begin{figure*}
    \includegraphics[width=\linewidth]{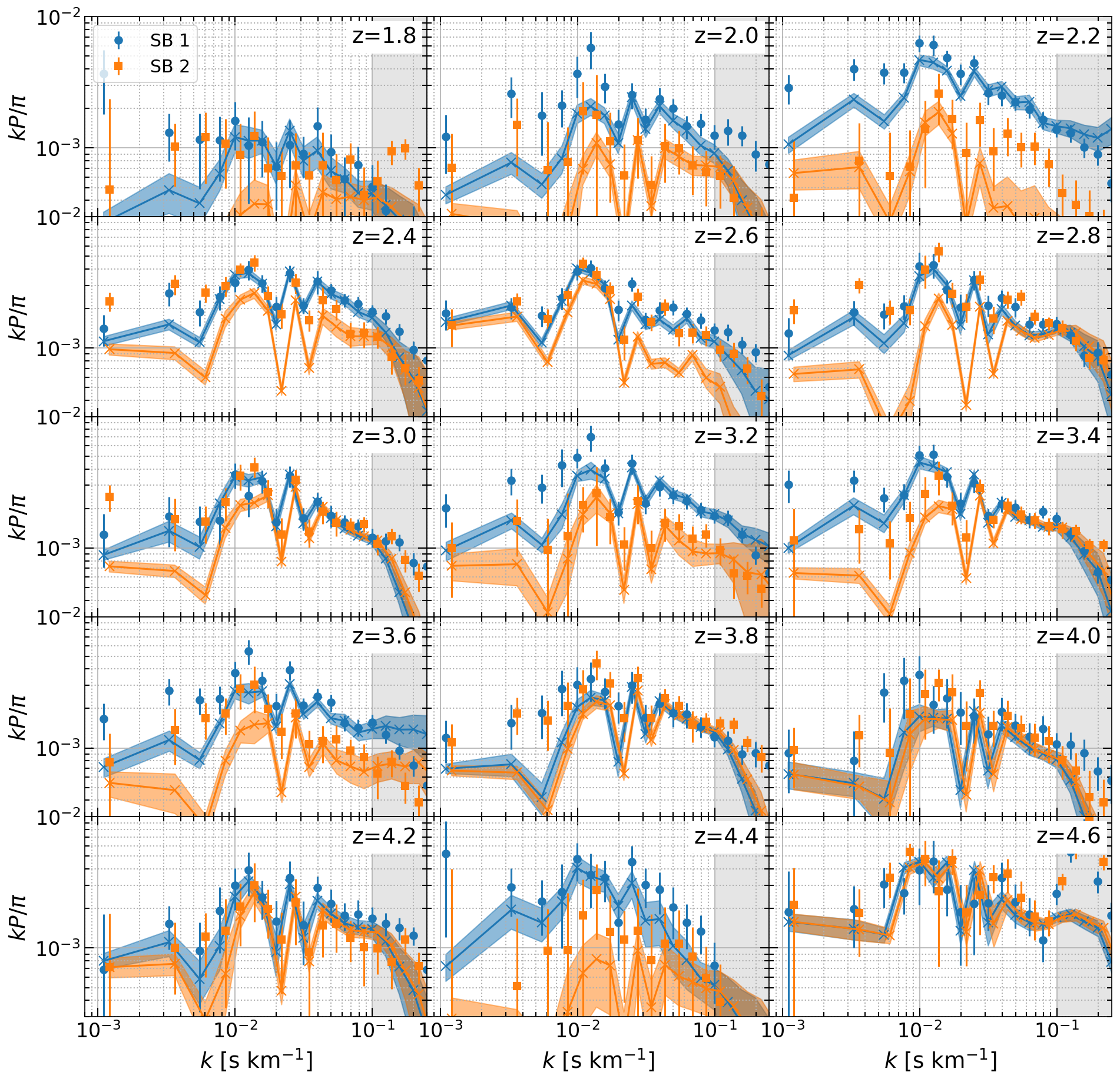}
    \caption{Data points (blue circles and orange squares) vs.\ our model (crosses with shaded regions) for high-resolution spectra.
    Here, three ions \ion{C}{IV}, \ion{Si}{IV} and \ion{Mg}{II} are modelled and the smooth power is not included.
    The mean power (solid lines with crosses) and its standard deviation (shaded regions) are calculated using all equally weighted posterior samples.
    We fit up to $k_\mathrm{max}=0.1$\skm (shaded grey region).
    }
    \label{fig:hr-compare-data-vs-model-mgii}
\end{figure*}

\subsection{Parameter inference}
We limit our parameter inference tests to three possible cases.
All three cases include parameters for \ion{C}{IV} and \ion{Si}{IV} at minimum.
We begin our tests with this minimum combination (referred as our baseline from now on).
We then introduce \ion{Mg}{II} to observe its effect on these estimations.
Note that the \ion{Mg}{II} doublet separation is close to the \ion{C}{IV} separation and therefore it is difficult to constrain using only the current two side bands.
Thus, we treat it as a systematic to be marginalized over.
Furthermore, our model does not account for large-scale clustering nor any error in the noise power subtraction.
To marginalize over these unmodelled effects, we introduce a simple additive smooth power $P_\mathrm{smooth}(k) = P_{sb}$, where $P_{sb}$ can be negative\footnote{To elaborate, we initially tried a scale dependence with a power law in our preliminary analysis.
However, we observed a strong degeneracy between amplitude and scale-dependence power, so we decided to remove the scale dependence.
We leave this to future work.}.
A negative amplitude would mean that the noise is over-subtracted.
We model each side band with different amplitudes.
The parameters for each case are listed in Table~\ref{tab:parameter-combos}.
Note that \ion{C}{IV}, \ion{Si}{IV}, \ion{Mg}{II} and $P_1$ parameters model the power in SB~1 ($P_{\mathrm{SB1}}$), whereas \ion{C}{IV}, \ion{Mg}{II} and $P_2$ parameters model $P_{\mathrm{SB2}}$.
As an explicit example, $P_{\mathrm{SB1}}$ and $P_{\mathrm{SB2}}$ for baseline + smooth case are given by:
\begin{align}
    P_{\mathrm{SB1}}(k) &= P_1 + \sum_{\mathclap{i \in \{\mathrm{C}~\textsc{iv}, \mathrm{Si}~\textsc{iv}\}} }10^{A_i} P_{1a}(k; b_i) + 10^{2A_i} P_{2a}(k; b_i), \\
    P_{\mathrm{SB2}}(k) &= P_2 + \sum_{\mathclap{i \in \{\mathrm{C}~\textsc{iv}\}}} 10^{A_i} P_{1a}(k; b_i) + 10^{2A_i} P_{2a}(k; b_i).
\end{align}

\begin{table}
    \centering
    \begin{tabular}{l|l}
        Label & Fitting parameters  \\
        \hline
        Baseline & $A_{\mathrm{C}~\textsc{iv}}, b_{\mathrm{C}~\textsc{iv}}, A_{\mathrm{Si}~\textsc{iv}}, b_{\mathrm{Si}~\textsc{iv}}$ \\
        Baseline + smooth & $A_{\mathrm{C}~\textsc{iv}}, b_{\mathrm{C}~\textsc{iv}}, A_{\mathrm{Si}~\textsc{iv}}, b_{\mathrm{Si}~\textsc{iv}}, P_1, P_2$ \\
        Baseline + \ion{Mg}{II} & $A_{\mathrm{C}~\textsc{iv}}, b_{\mathrm{C}~\textsc{iv}}, A_{\mathrm{Si}~\textsc{iv}}, b_{\mathrm{Si}~\textsc{iv}}, A_{\mathrm{Mg}~\textsc{ii}}, b_{\mathrm{Mg}~\textsc{ii}}$ \\
        \hline
    \end{tabular}
    \caption{Fitting parameters for three combinations we test in this work.
    We fit each redshift bin independently.}
    \label{tab:parameter-combos}
\end{table}

To infer the cosmological parameters, we use the \textsc{ultranest}\footnote{\url{https://johannesbuchner.github.io/UltraNest}} \citep{ultranest} software, which is especially easy to setup and use.
More importantly, the underlying nested sampler algorithm is able to handle many difficult problems.
Nested-sampling has become a common tool in many fields since its inception \citep{skillingNestedSampling2004, skillingNestedSamplingGeneral2006}.
We refer the reader to \citet{ashtonNestedSamplingPhysical2022} for a nice review that focused on the physical sciences.

To speed up calculations, we cache interpolation points for the Doppler parameter $b$.
The lowest and highest value of $b$ also constitute a hard prior for each ion.
We calculate the template power spectra on a fine grid with velocity spacing $dv=1$\kms and $2^{16}$ points using FFTs.
The integrations over column densities are calculated as discrete rectangles of $d\log_{10}N=0.1$ size ranging from $\log_{10}N_1=11$ to $\log_{10}N_2=16$.
We fit each redshift bin independently and all side bands simultaneously (note again that \ion{Si}{IV} will not add power to SB~2).

The power spectrum is measured in discrete $k$ bins of a certain width and this binned measurement is the integral average of a model within that bin.
Furthermore, our quadratic estimator actually measures the power spectrum as deviations from a fiducial power as discussed in the previous section, which alleviates the averaging effect in bins.
In our $\chi^2$ and likelihood calculations, 
we take this effect into account by summing model power minus fiducial at $n_\mathrm{sub}$ discrete points linearly spaced within a bin.

Various challenges along the way pointed us towards a nested sampler.
Due to a combination of data and modelling difficulties, the biggest challenge has been the instability.
For example, the signal is small such that any contamination in the data will have relatively large impact.
These contaminants could come from continuum errors at large scales, resolution errors at small scales, calibration errors, sky subtraction errors etc.
Fitting the power spectrum is not ideal either, as the model in Figure~\ref{fig:model_fid} shows the blending of many features.
Furthermore, any unmodeled ions or clustering will affect all scales.
Therefore, our work requires a method and software that is robust against this multifaceted numerical challenge.
We initially used the minimizer \textsc{iminuit}\footnote{\url{https://iminuit.readthedocs.io}}
\citep{iminuit}, and then sampled around the minimum using \textsc{emcee}\footnote{\url{https://emcee.readthedocs.io}} sampler \citep{emcee}.
However, neither were robust against these problems.
The minimizer often failed to produce a valid fit and the sampler got stuck around local minima, and did not converge.
Our most successful tool is the nested sampler.

\section{Data and analysis}
\label{sec:data_analysis}
To assess the descriptiveness of our model and challenges regarding parameter inference and systematics in data,
we apply our model to real side band power spectrum measurements.
In the first subsection, we analyze publicly available, high-resolution quasar spectra and in the second subsection we analyze early data from DESI. 
This data analysis and parameter inference are mainly exploratory, and we focus on \ion{C}{IV} while marginalizing over \ion{Si}{IV} and \ion{Mg}{II}.

\subsection{High-resolution spectra}
We use the 1D side band power spectra from \cite{karacayliOptimal1DLy2022}.
That work used three publicly available data sets with high-resolution quasar spectra to measure the Ly~$\alpha$ power spectrum. 
They find clear doublet features in both the power spectrum and correlation function in the side bands.
Of the three datasets, only two were used for the side-band (SB) power spectrum estimation and are therefore analyzed here:
\begin{itemize}
    \item Keck Observatory Database of Ionized Absorption toward Quasars (KODIAQ) Data Release 2\footnote{\url{https://koa.ipac.caltech.edu/workspace/TMP_939bFW_53591/kodiaq53591.html}} (DR2) \citep{lehnerGALACTICCIRCUMGALACTICVI2014,omearaFIRSTDATARELEASE2015,omearaSecondDataRelease2017} from observations with HIRES   \citep{vogtHIRESHighresolutionEchelle1994} on the Keck I telescope. 
    This dataset has 300 reduced, continuum-fitted, high-resolution quasar spectra at $0<z<5.3$ with resolving power $R \gtrsim 36\,000$.
    The continuum is fitted by hand one echelle order at a time using Legendre polynomials.

    \item The Spectral Quasar Absorption Database (SQUAD) DR1\footnote{\url{https://doi.org/10.5281/zenodo.1345974}} \citep{murphyUVESSpectralQuasar2019} from observations with UVES \citep{dekkerDesignConstructionPerformance2000} on the European Southern Observatory’s Very Large Telescope (VLT). 
    This dataset consists of 467 fully reduced, continuum-fitted, high-resolution quasar spectra at redshifts $0<z<5$ with resolving power $R \gtrsim$ 40\,000. 
    Its continuum fitting consists of an automatic phase and then a manual phase to eliminate the remaining artifacts. 
\end{itemize}

\begin{figure}
    \centering
    \includegraphics[width=\columnwidth]{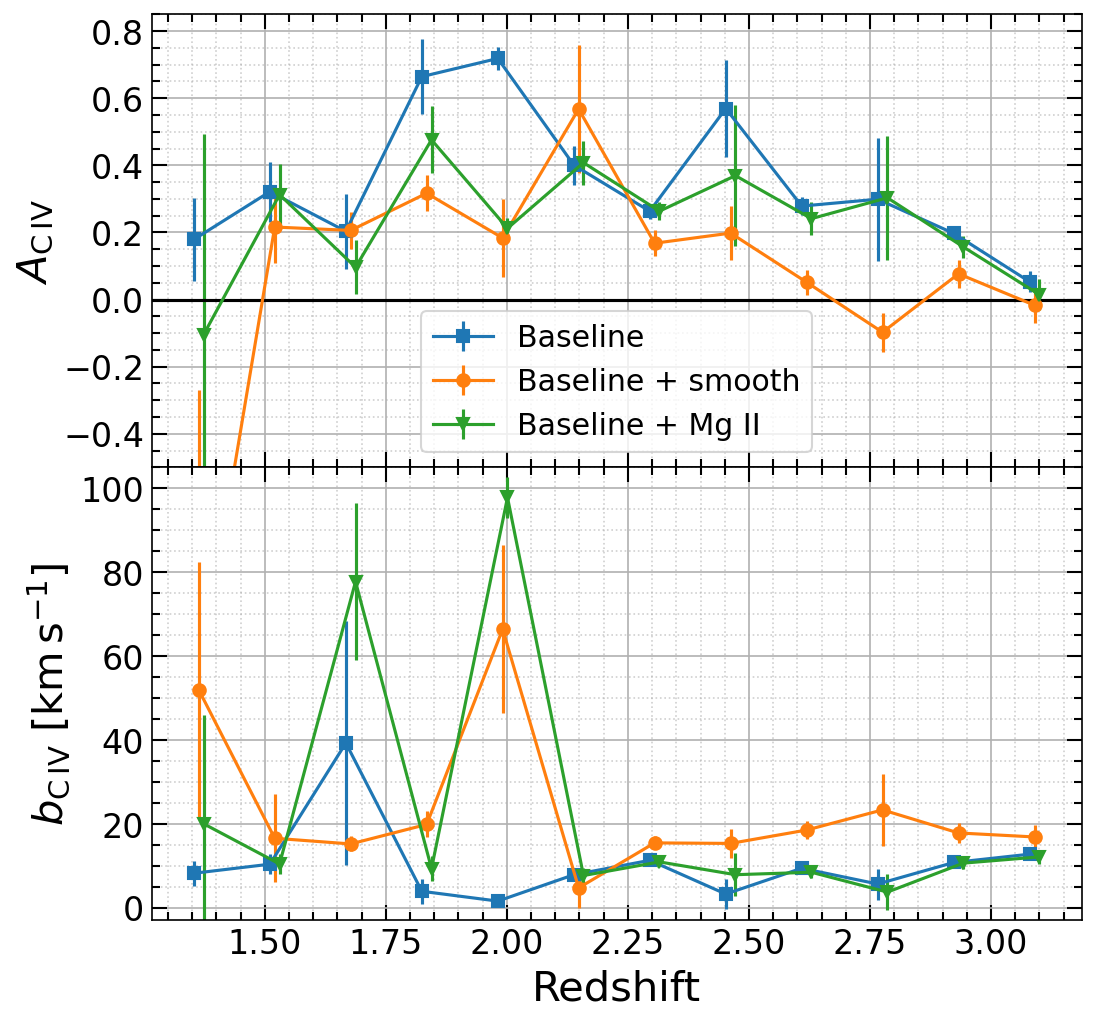}
    \includegraphics[width=\columnwidth]{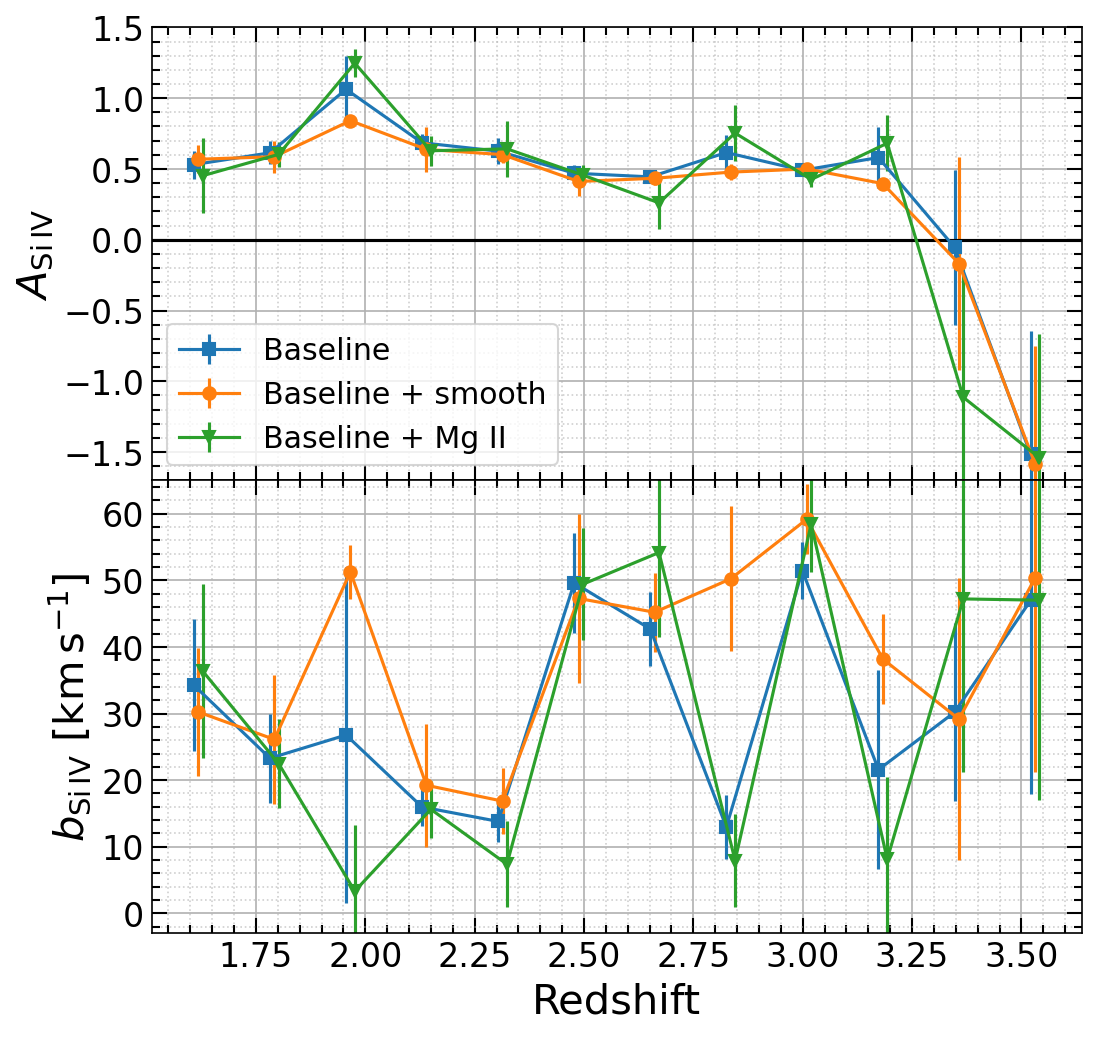}
    \caption{Amplitude and Doppler parameter from high-resolution analysis for \ion{C}{IV} (top panel) and \ion{Si}{IV} (bottom panel) as a function of redshift.
    The three sets of points and lines correspond to our baseline model (blue squares and lines), the baseline model plus a smooth component (orange circles and lines), and the baseline model plus \ion{Mg}{II}. Redshift is with respect to the absorber transition.}
    \label{fig:hr-civ-siv-compare-ab}
\end{figure}

\begin{figure}
    \centering
    \includegraphics[width=\columnwidth]{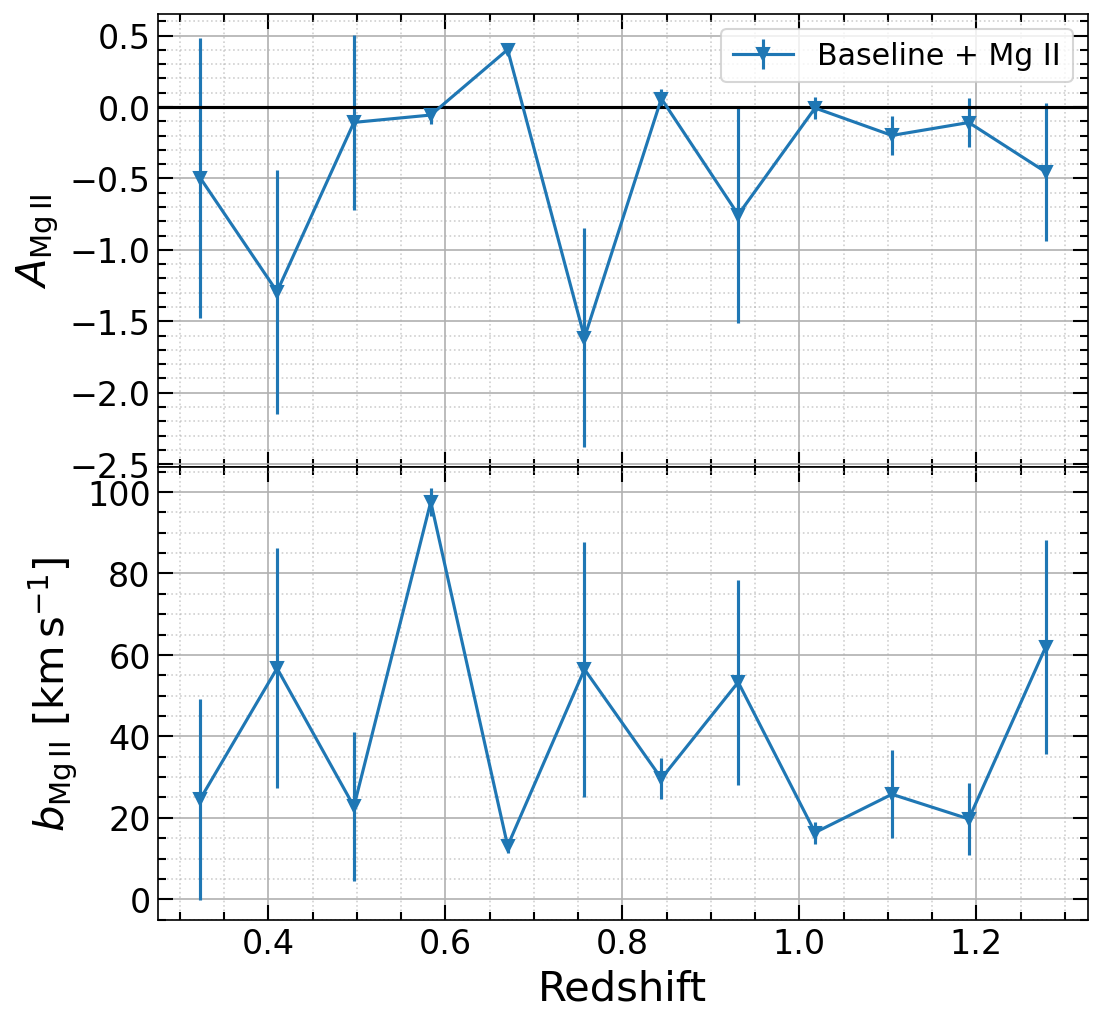}
    \caption{Amplitude and Doppler parameter from high-resolution analysis for \ion{Mg}{II} as a function of redshift.
    The signal most likely comes from high-column density systems, i.e. nearly saturated lines.
    The results are broadly consistent with our fiducial value of $A=0$.}
    \label{fig:hr-mg-ab}
\end{figure}


\cite{karacayliOptimal1DLy2022} further re-samples all these high-resolution spectra  onto a common 3~km\,s$^{-1}$ grid. 
To measure the power spectrum, they apply the optimal quadratic estimator, which is robust against spectral masking and gaps, and deconvolve the spectrograph window function from the results.
They use the following fiducial power spectrum parameters in equation~(\ref{eq:pd13_fitting_fn}): $A=2.71\times 10^{-3}$, $n=-2.92$, $\alpha=-0.174$, $B=2.36$ and $\beta=-0.014$.
Note that the Lorentzian term is zero for side band power.
They devise a regularized bootstrap method to obtain the statistical covariance matrix.
In this work, we use the full covariance matrix.
We add the resolution systematic errors by rescaling the reported budget with respect to the fiducial power used in the SB estimation.
Continuum errors and their systematics budget can be ignored in the side bands, since this region of the quasar spectrum is absorption-free and the continuum can be extracted faithfully.

\begin{figure*}
    \includegraphics[width=\linewidth]{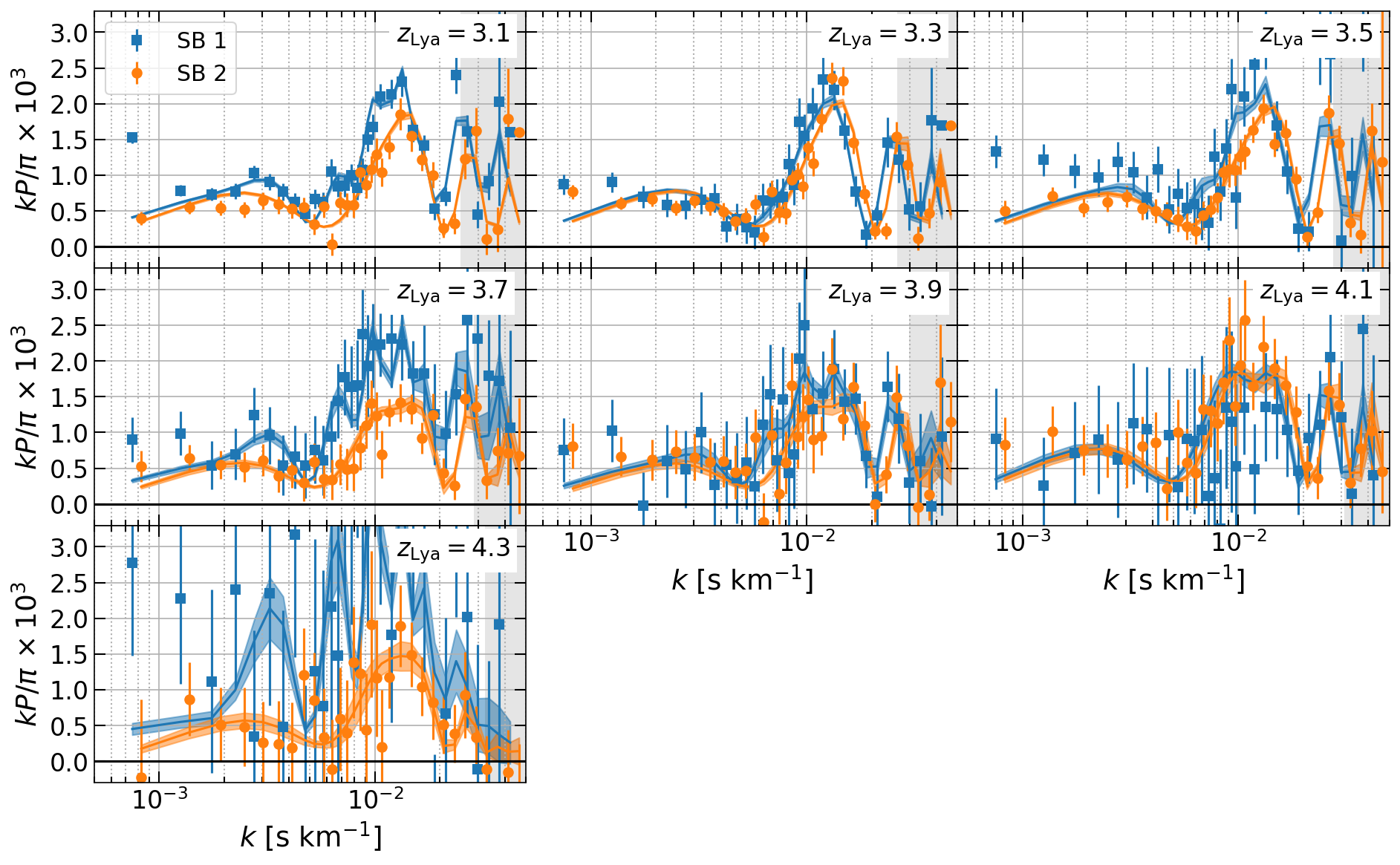}
    \caption{Data points vs.\ model fits for DESI analysis.  DLA and sub-DLA sightlines have been removed.
    Error bars on data points are from 5\,000 bootstrap realizations of 320 subsamples.
    Gaussian values are replaced only if they are smaller than bootstrap estimates.
    We model three ions (\ion{C}{IV}, \ion{Si}{IV} and \ion{Mg}{II}), and leave out smooth power.
    The mean power (solid lines with crosses) and its standard deviation (shaded regions) are calculated using all equally weighted posterior samples.
    Our confidence regions are $0.5\times 10^{-3}<k<1.2/R_\mathrm{kms}$.
    }
    \label{fig:fugu-sb-power-nondla-vs-all}
\end{figure*}

We create and cache 21 interpolation points of Doppler parameter $b$ between [1, 101]\kms.
We assume a hard prior of [-3, 3] for $A_\mathrm{ion}$ (deviation of $\log_{10} f(N)$ amplitude) for all ions.

Figure~\ref{fig:hr-compare-data-vs-model-mgii} shows data points and the model power spectrum when \ion{Mg}{II} is taken into account.
We calculate all model power spectra using all equally weighted posterior samples.
The line corresponds to the mean of this power, while shaded regions are the standard deviations.

The \ion{C}{IV} peak is the most visible feature in data.
However, other ion features could be present and introduce uncertainties to the measurement.
Figure~\ref{fig:hr-civ-siv-compare-ab} shows the amplitude and Doppler parameter for \ion{C}{IV} (top panel) and \ion{Si}{IV} (bottom panel) for the baseline model, and the baseline model with the addition of \ion{Mg}{II}, and the baseline model with the addition of a smooth component $P_{SB}$.
In this figure, redshift is with respect to the absorber line and the data points are obtained from the mean and standard deviation.
One important caveat is that strong metal lines from high-column density systems are not masked in the SB power estimation.
This means our abundance and Doppler parameter measurements do not only represent IGM metals.
Adding contaminants to \ion{C}{IV} lowers the measured amplitude and increases the Doppler parameter.
Even though the \ion{Si}{IV} results do not seem to be affected by the contaminants, the measured $b$ values are too large to be physical. 
Moreover, the results seem most unstable below $z_\mathrm{\ion{C}{IV}} \lesssim 2.0$, which manifests as strong jumps and increased error bars on the Doppler parameters, and some redshifts yield bimodal results.
We adopt physically motivated values with \ion{Mg}{II} contamination as our main result.

We checked the amplitude and Doppler parameter for \ion{Mg}{II} for the baseline model plus \ion{Mg}{II} and show the results in Figure~\ref{fig:hr-compare-data-vs-model-mgii}. Even though the signal is most likely dominated by high-column density systems, such as \ion{Mg}{II} systems that reside in damped Ly~$\alpha$ absorbers, the results appear broadly consistent with our fiducial value of $A=0$.

\begin{figure}
    \centering
    \includegraphics[width=\columnwidth]{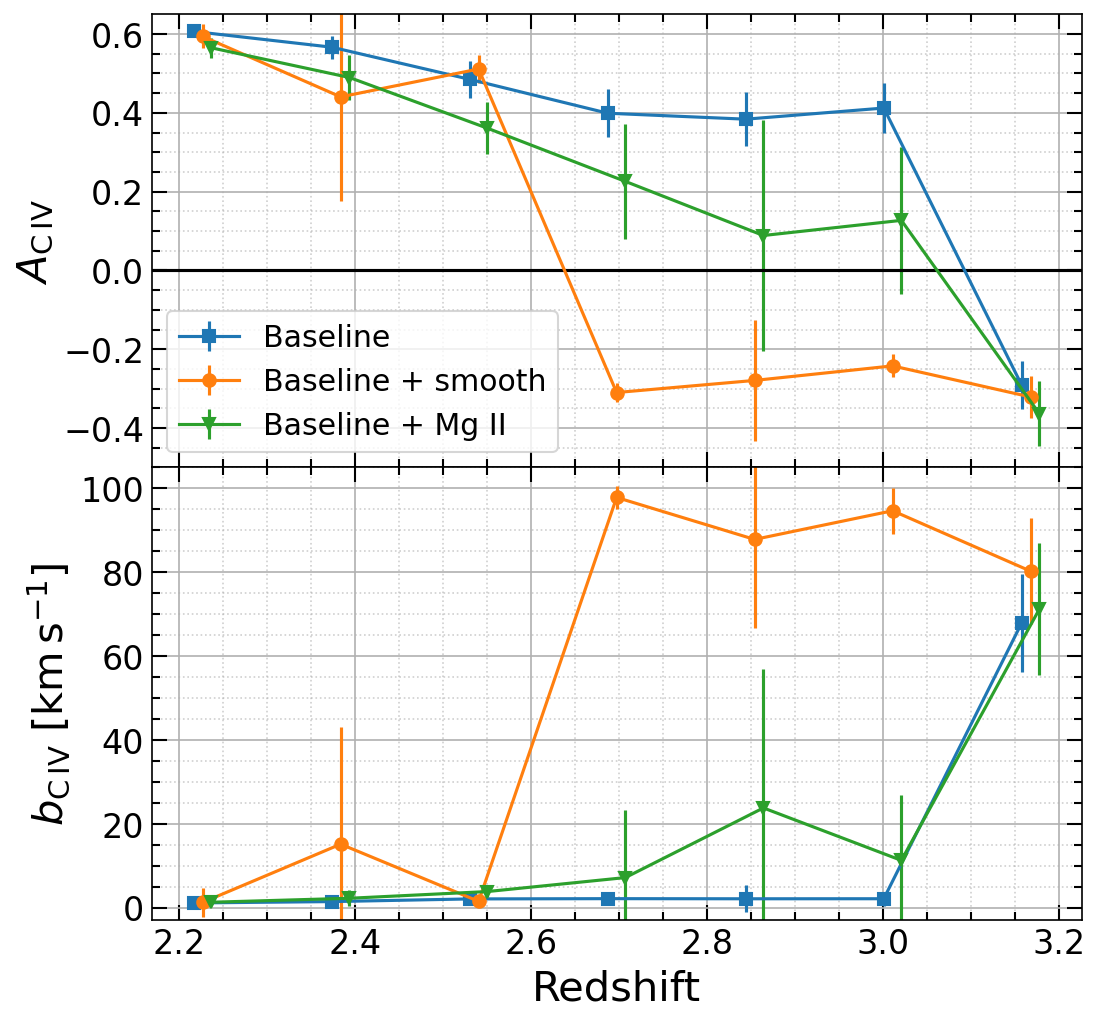}
    \includegraphics[width=\columnwidth]{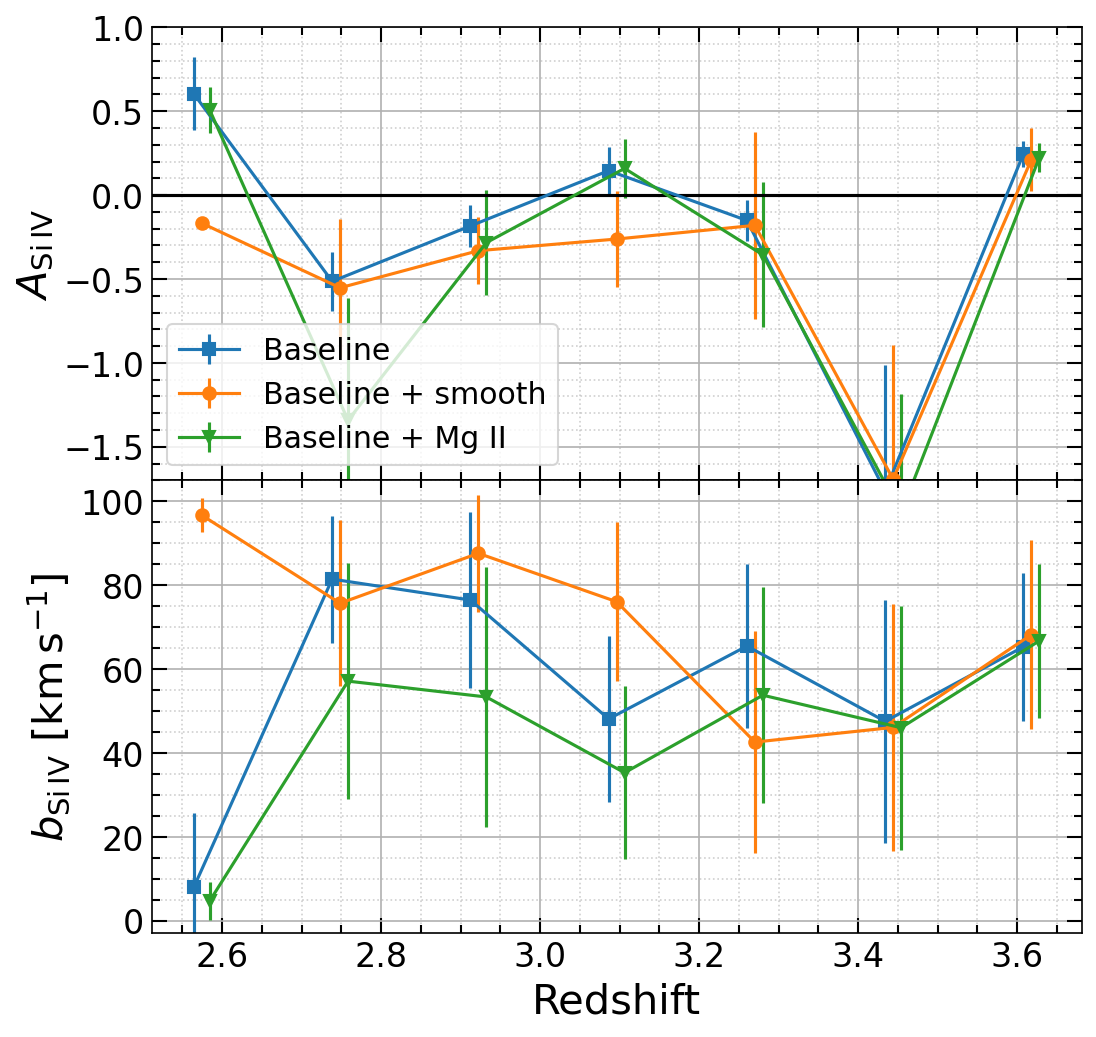}
    \caption{Amplitude and Doppler parameter from DESI analysis when DLA sightlines are removed against these two contaminants for \ion{C}{IV} (top panel) and \ion{Si}{IV} (bottom panel).
    Redshift is with respect to the absorber transition.}
    \label{fig:fugu-civ-siv-compare-ab}
\end{figure}

\subsection{DESI early data}
\label{subsec:desi_edr}
A major strength of our model is its capability to exploit large numbers of noisy spectra.
DESI can observe nearly 5\,000 objects simultaneously in a single exposure \citep{desicollaborationDESIExperimentPart2016b, silberRoboticMultiobjectFocal2023}, and
is currently conducting a five-year survey to observe millions of quasars with moderate SNR and improved resolution over SDSS \citep{desicollaborationDESIExperimentPart2016}.
Therefore, DESI has access to higher $k$ modes in the power spectrum, which are needed to better constrain the $b$ parameter in our model.

DESI has conducted a Survey Validation (SV) period to test various target selection methods before it began the five-year main survey in May 2021.
The first phase of SV collected deep observations that were used to optimize the selection algorithms with a visually inspected sample \citep{chaussidonTargetSelectionDESIQSO2022, alexanderDESISVVIQSO2022}. 
Another phase was aimed at studying the clustering program and covered about 1\% of the DESI main survey.
We use these two surveys that are part of early data release (EDR), and further include two months of main survey to increase the statistical precision in our analysis.
We limit ourselves to objects that are targeted as quasars in the afterburner catalogue \citep{deyOverviewDESILegacy2019, yechePreliminaryTargetQSODESI2020, chaussidonTargetSelectionDESIQSO2022, alexanderDESISVVIQSO2022}.

The continuum fitting algorithm we use has been developed over the last few years and applied to many Ly~$\alpha$ forest BAO analyses \citep{bautistajuliane.MeasurementBaryonAcoustic2017, bourbouxExtendedBaryonOscillation2019, bourbouxCompletedSDSSIVExtended2020}.
We model every quasar continuum $\overline{F} C_q(\lambda_\mathrm{RF})$ with a global mean continuum $\overline{C}(\lambda_\mathrm{RF})$ and two quasar "diversity" parameters, an amplitude $a_q$ and slope $b_q$:
\begin{align}
    \overline{F}C_q(\lambda_\mathrm{RF}) &= \overline{C}(\lambda_\mathrm{RF}) \left( a_q + b_q \Lambda \right) \\
    \Lambda &= \frac{\log\lambda_\mathrm{RF} - \log\lambda_\mathrm{RF}^{(1)}}{\log\lambda_\mathrm{RF}^{(2)} - \log\lambda_\mathrm{RF}^{(1)}}
\end{align}
where $\lambda_\mathrm{RF}$ is the wavelength in the quasar rest frame.
Note our quasar continuum definition absorbs the IGM mean flux $\overline{F}(z)$ into $a_q$ and $b_q$.
The software Package for Igm Cosmological-Correlations Analyses (\textsc{picca}) is publicly available\footnote{\url{https://github.com/igmhub/picca}}.
It fits every quasar for $a_q$ and $b_q$, and stacks the residuals in the rest-frame to update the global mean continuum $\overline{C}(\lambda_\mathrm{RF})$.
To effectively study the effects of DLAs, we demand all of Ly~$\alpha$ forest region to be present in spectra.
Given the minimum DESI observed wavelength is 3600\,\AA\ and taking the forest lower end to be 1040\,\AA\ in the rest-frame results in a quasar redshift cut of $z_\mathrm{qso}>2.43$.
We define SB~1 (\ion{Si}{IV} forest) to be between 1268--1380\,\AA\ and SB~2 (\ion{C}{IV} forest) to be between 1409--1523\,\AA\ in the rest-frame.
Given our quasar redshift cut, SB~1 starts at 4350\,\AA\ and SB~2 starts at 4830\,\AA\ in the observed frame.
We further demand all spectra to observe SB~2.
We use \textsc{picca} to calculate the continuum multiplied by the mean flux $\overline{F}C_q(\lambda_\mathrm{RF})$ in 4830--6500\,\AA\ in the observed wavelength range and in $\Delta\lambda_{\mathrm{RF}}=2.5$\,\AA\ coarse rest-frame binning pixels.
Here we require SNR greater than 0.25 to weed out possible contaminants.
We mask BAL features in the spectrum using both the absorptivity (AI) and balnicity indices (BI) criteria \citep[e.g., see ][]{guoClassificationBALs2019, ennesserMitigationBALquasars2022}.
We also mask major sky lines\footnote{\url{https://github.com/corentinravoux/p1desi/blob/main/etc/skylines/list_mask_p1d_DESI_EDR.txt}} and do not consider spectra with less than 50 of the 2.5\,\AA\ wide pixels.

After all these masks and cuts, we are left with 41\,341 quasars with data for SB~1 and 48\,497 quasars in SB~2.
Removing sightlines that have DLAs and sub-DLAs (i.e. $\log N_\mathrm{H\textsc{i}}>19$) reduces these numbers to 11\,395 and 15\,346 \citep{wangDeepLearningDESIDLA2022}.
These numbers are listed in Table~\ref{tab:sb_numbers}.
For brevity, we present results on non-DLA and non-sub-DLA quasars unless explicitly stated.

\begin{table}
    \centering
    \begin{tabular}{l|c|c|c}
         & Wavelength range [\AA] & \# All quasars & \# Non-DLA quasars \\
        \hline
        SB~1 & 1268--1380 & 41\,341 & 11\,395 \\
        SB~2 & 1409--1523 & 48\,497 & 15\,346 \\
        \hline
    \end{tabular}
    \caption{Side band rest-frame wavelength ranges and number of quasars in DESI early data.
    For our default analysis we also remove sightlines that have DLAs and sub-DLAs (non-DLA quasars).}
    \label{tab:sb_numbers}
\end{table}


The quadratic estimator splits the spectra into two if they have more than 500 pixels.
We are using the resolution matrix produced by the spectroscopic pipeline based on tests on pixel-level simulations (Kara{\c c}ayl{\i} et al., in preparation).
We smooth the noise estimates in the covariance matrix by a hybrid Gaussian box-car window function.
The box size is 50 pixels with a Gaussian sigma of 20 pixels.
We marginalize out three modes of continuum errors: constant, $\ln \lambda$ and $(\ln \lambda)^2$ polynomials.
The fiducial power spectrum has $A=2.084\times 10^{-3}$, $n=-3.075$, $\alpha=-0.07423$, $B=1.599$ and $\beta=-0.2384$.
Finally, we measure the power spectrum in 20 linear, 20 log-linear $k$ bins with $\Delta k_\mathrm{lin} = 0.0005$\skm and $\Delta k_\mathrm{log} = 0.05$ bin sizes in 7 redshift bins from 3.1 to 4.3 with respect to Ly~$\alpha$ transition line.
We perform only one iteration.
Subsequent iterations mostly refine Fisher matrix estimates, which we replace with bootstrap analysis \citep{karacayliOptimal1DLy2020}.

We generate 5\,000 bootstrap realizations over 320 subsamples for each side band.
Even though our implementation of QMLE generates correlations between redshift bins, we ignore all correlations between power spectrum bins, and assume the covariance matrix is diagonal.
We replace Gaussian errors only if they are smaller than bootstrap estimates.

Figure~\ref{fig:fugu-sb-power-nondla-vs-all} shows our power spectrum measurement in these two side bands, where the 
error bars are from bootstrap realizations.
We report the redshift according to the Ly~$\alpha$ transition line.
Each ion will be at a different redshift.
The oscillations produced by \ion{C}{IV} are clearly visible at all redshifts.

We perform nested sampling with same settings as high-resolution analysis.
Figure~\ref{fig:fugu-civ-siv-compare-ab} shows the $A$ and $b$ parameters for \ion{C}{IV} in the top panel and \ion{Si}{IV} in the bottom panel.
The most stable results come from \ion{C}{IV} as expected.
The abundance $A$ seems to be higher compared to our fiducial measurement from \citet{scannapiecoSourcesIntergalacticMetals2006}, and shows a downward trend over redshift (note even though our resolution improves at higher redshifts (wavelengths), our statistics decline.).
The Doppler parameter $b$ mostly remains under 20\kms; and any seemingly nonphysical values and jumps are only present with error bars.
Marginalizing over a smooth power component adds sudden jumps in the results, which could mean it is a highly degenerate nuisance parameter.
Introducing \ion{Mg}{II} (a more physically-motivated nuisance model) lowers the amplitduce $A_{\mathrm{C}~\textsc{iv}}$ and increases the uncertainties in a more stable fashion.
As previously, we take the results with \ion{Mg}{II} as our main results.
However, the \ion{Mg}{II} results themselves are not trustworthy as the $b$ value consistently hits the upper boundary of its prior of 100\kms.

\subsection{Comparison between DESI and high-resolution}
\begin{figure}
    \centering
    \includegraphics[width=\columnwidth]{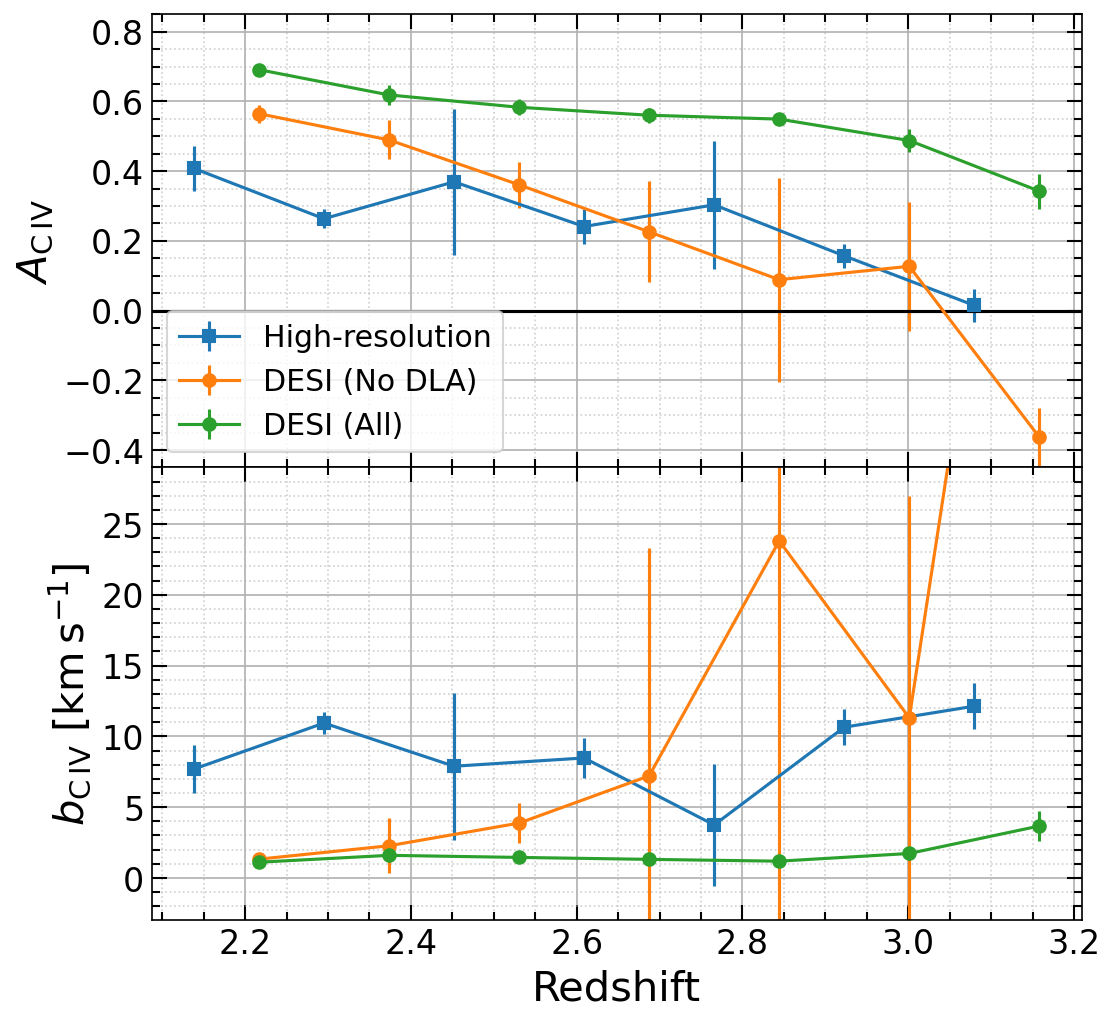}
    \caption{Comparison of $A$ and $b$ values of \ion{C}{IV} measurements between DESI and high-resolution spectra. The No DLA case excludes both DLAs and sub-DLAs. 
    All three datasets all show a clear downward trend with increasing redshift.}
    \label{fig:fugu-vs-highres-civ}
\end{figure}

Figure~\ref{fig:fugu-vs-highres-civ} compares our $A$ and $b$ measurements for \ion{C}{IV} for the high resolution data to the DESI data both with and without DLAs.
The effects of \ion{Si}{IV} and \ion{Mg}{II} are marginalized over.
Interestingly enough, high-resolution \poned (with DLA sight-lines) agrees well with the DESI non-DLA measurement.
All three measurements show a clear downward trend with redshift similar to \citet{yangMetalLinesAssociated2022}, indicating a gradual increase in the abundance of these metals at $z\simeq 3.0$.
However, as we discuss in Section~\ref{sec:discussion}, our findings could be partially attributed to the growth of cloud-cloud clustering.

\section{Discussion}
\label{sec:discussion}
\textbf{Correlation function vs.\ power spectrum:}
The reader may already suspect that this analysis would have been more robust if it were performed on the correlation function instead.
After all, the features are localized as peaks in correlation function, whereas they are modulated and superposed in the power spectrum.
The challenge for correlation function comes in two major points.
First, the spectrograph resolution smooths the correlation function at all scales.
Without a quadratic estimator for $\xi_F$, we would have had to convolve the model power spectrum with an average window function, similar to 3D analyses.
We are spared from this complication because our quadratic estimator deconvolves the individual, wavelength-dependent spectrograph resolution from each spectrum based on spectro-perfectionist extraction \citep{boltonSpectroPerfectionismAlgorithmicFramework2010}.
Second, the continuum errors distort the correlation function at all scales.
Again, without a quadratic estimator for $\xi_F$, a distortion matrix has to be introduced into the model similar to what is done for 3D studies.
These distortions are localized to large-scale modes (low $k$) in power spectrum, and are marginalized out by our quadratic estimator.
Therefore, the correlation function analysis would be more robust only with a quadratic estimator.

\textbf{Metal contamination in the Ly~$\alpha$ forest:}
These same metals are also present in the Ly~$\alpha$ forest, and are considered contaminants in the Ly~$\alpha$ \poned analyses.
Because of strong \ion{H}{I} absorption, metal lines cannot be identified even in high-SNR spectra in the Ly~$\alpha$ forest, and therefore must be removed statistically from the Ly~$\alpha$ power spectrum.
This is achieved by subtracting the estimated SB~1 power from the Ly~$\alpha$ forest estimates \citep{mcdonaldLyUpalphaForest2006, palanque-delabrouilleOnedimensionalLyalphaForest2013, chabanierOnedimensionalPowerSpectrum2019, karacayliOptimal1DLy2022}.
However, this only removes power due to metals with $\lambda_{\mathrm{RF}} \gtrsim 1400$\,\AA,
and hence leaves behind some metal contamination such as \ion{Si}{III}, which produces oscillatory features due to the \ion{Si}{III}-Ly\,$\alpha$ cross correlation \citep{mcdonaldLyUpalphaForest2006, palanque-delabrouilleOnedimensionalLyalphaForest2013}.
Our model can potentially better remove these $\lambda_{\mathrm{RF}} \lesssim 1400$\,\AA\ metals from the Ly~$\alpha$ \poned.
Such an application requires also including singlets into the model, which unfavourably do not manifest any characteristic features in the two-point statistics.

\textbf{Growth of cloud-cloud clustering:}
We assume cloud-cloud clustering $\xi_{cc}$ is constant with respect to redshift.
However, there are hints for structure growth in $\xi_{cc}$ \citep{scannapiecoSourcesIntergalacticMetals2006}.
Even though this term only is important for the 2a term, we showed that the 2a term is usually larger than the 1a term.
Therefore, some or even all of the redshift trend for the $A$ (abundance) parameter could be due to enhanced clustering of metals.
A more proper analysis would reformulate equation~(\ref{eq:fitting_xiall}):
\begin{align}
    A_\mathrm{ion}(z) &= A_0 + A_1 \log\left(\frac{1+z}{1+z_0}\right) \label{eq:discuss:a_ion_z} \\
    D_\mathrm{ion}(z) &= D_0 + D_1 \log\left(\frac{1+z}{1+z_0}\right), \label{eq:discuss:D_ion_z}
\end{align}
and then fit for all redshifts simultaneously.
This could potentially produce tighter constraints on modelled parameters, but requires better understanding of the wavelength dependence of systematics and related correlations between redshift bins.
We leave this to a future study.

\textbf{Large-scale cloud-cloud clustering:}
In this work, we empirically quantify cloud-cloud clustering using direct observations.
The data and our formulation work better on small scales, and cannot properly capture large-scale correlations, which may become important for the 2a term at low $k$.
One might try the standard tracer-biased power spectrum:
\begin{equation}
    P_\mathrm{ion}(\bm k) = b^2_\mathrm{ion} (1+\beta_\mathrm{ion}\mu^2)^2 P_L(k) F(\bm k),
\end{equation}
where $b_\mathrm{ion}$ is the bias, $\beta_\mathrm{ion}$ the redshift space distortion parameter, $P_L$ is the linear power spectrum, and $F(\bm k)$ accounts for non-linear effects \citep{bourbouxCompletedSDSSIVExtended2020}.
However, since these metal systems are heavily governed by small-scale physics, it is unclear if this modelling would be advantageous over an empirical formulation.

\textbf{Thermal broadening of \ion{C}{IV}:}
At the expected IGM temperature of 15\,000~K, the carbon atom has $b_\mathrm{\ion{C}{IV}}=4.5$\kms from pure thermal broadening.
We find somewhat higher but consistent results.
Higher temperatures and/or turbulent motion can physically explain our $b_\mathrm{\ion{C}{IV}}$ findings.
For example, these systems may not reside in the IGM despite our efforts to eliminate such sources through the exclusion of sightlines with DLAs and sub-DLAs.
A possible candidate is the strong and blended Ly~$\alpha$ forest systems where the column densities are not high enough to self shield and show different metal properties \citep{perezrafolsGalaxiesinAbsorption2022}.
However, systematic errors including resolution effects and model fitting deficiencies are more likely explanations.
Therefore, we reserve our temperature conclusions for a future study.

\textbf{Line blending:}
Our model 2a term breaks down when lines get saturated at high column densities such that they cannot be superposed.
This overlapping happens at scales of $b\sim 10$\kms or the corresponding spectrograph resolution, whichever is higher;
and it is more important for ions with smaller correlation lengths $r_0$ (see table~\ref{tab:fiducial_fN_xicc}).
This effect is noted as the absorber exclusion in \citet{irsicAbsorberModelHalolike2018}, which is again analogous to the halo exclusion in the halo model \citep{coorayHaloModelsLarge2002}.
However, the absorption profiles are spread out due to temperature broadening or the spectrograph resolution, and are free to overlap in flux.
The problem occurs when the system is saturated such that $K=1-\mathrm{e}^{-\tau} \approx \tau$ is no longer valid.
This exclusion effect might be a bigger problem for \ion{H}{I},
but should be less concerning for weak metal lines.

\textbf{Addition of other side bands:}
Our model is also capable of constraining \ion{Mg}{II} values,
but results from this analysis are not particularly reliable.
Ideally, we would isolate the \ion{Mg}{II} signal in a "third" side band.
A common choice is 1600–-1800\,\AA\ in the rest-frame \citep{bourbouxExtendedBaryonOscillation2019}.
In principle, one can consider as many side-bands as one desires, but pipeline noise systematics will eventually become dominant.
This might be true even with just SB~3 of the current data set since the \ion{Mg}{II} signal is weak.
Generally, systematics will make analyses for weak signals including \ion{Mg}{II} especially difficult.

\textbf{Correlations between different species:}
One might be tempted to include correlations between different ions (e.g. \ion{C}{IV}-\ion{Si}{IV}).
However, we must be aware of the side band lengths before modeling such correlations.
SB~1 is about 11\,000\kms and SB~2 is 10\,000\kms in length, 
whereas the velocity separation between the \ion{C}{IV} and \ion{Si}{IV} transitions is 14\,000\kms.
Similarly, many transition lines between different species occur at large separations.
Therefore, the cross-correlations between the most conspicuous ions will not be present in data at all.
The study of possible combinations is out of the scope of our work.

\textbf{\ion{Fe}{II} specific complication:}
All three ions we considered in this work show only one doublet feature, and therefore their contributions to one Ly~$\alpha$ redshift bin come from single (separate) sources.
In contrast, \ion{Fe}{II} shows two main doublet features.
This means two \ion{Fe}{II} sources at different redshifts contribute to the same Ly~$\alpha$ redshift bin.
This interconnection of multiple redshifts are better handled in an analysis using equations~(\ref{eq:discuss:a_ion_z}) and (\ref{eq:discuss:D_ion_z}).

\textbf{Recovering DLA sight-lines:}
For simplicity, we removed all spectra that show high-column density systems.
This conservative choice significantly reduces the SNR and constraining power of the data set.
In a future study, we plan to only mask regions that correspond to all possible metal transition lines, which should add more quasar spectra to the analysis.

\textbf{Studying metals in CGM or high-column density systems:}
Our model can also be extended to studying metals in the CGM or high-column density systems instead of removing them.
For example, we can simply introduce another population into the column density distribution $f(N)$:
\begin{equation}
    f(N) = f_\mathrm{IGM}(N) + f_\mathrm{CGM}(N)
\end{equation}
Ideally, cross-correlations between the two populations would be inconsequential.

\textbf{High column density cut-off:}
As \citet{cookseyPreciousMetalsSDSS2013} note, a power law for $f(N)\propto N^{-\alpha}$ formally corresponds to infinite $\Omega_\mathrm{\ion{C}{IV}}$ for $\alpha<2$.
\begin{equation}
    \Omega_{\ion{C}{IV}} = \frac{H_0 m_C}{c \rho_{c,0}} \int f(N) N \ddif N \propto N^{2-\alpha}
\end{equation}
Studies such as \citet{scannapiecoSourcesIntergalacticMetals2006} and ours usually limit the integration ranges.
As we have shown in Section~\ref{subsec:dependence_on_n}, divergence is not a problem for our model since $\log N>16$ systems contribute little \poned.
However, a power law $f(N)$ is not physical either, so we expect a high column density cut-off $N_\mathrm{cut}$ in $f(N)$.
Furthermore, if $\log N_\mathrm{cut} \approx 16$, our method can constrain this value with better data.

\section{Summary}
\label{sec:summary}
The metal abundance in the universe and its evolution with redshift has implications for the enrichment and clustering of the CGM and IGM.
There is a rich literature that individually detects metals lines in quasar spectra to study the properties of that sample.
These studies connect local environments to the metal systems, but come with possible selection biases from SNR limited samples and incompleteness issues due to identifier efficiency for a cosmological measurement.

In this work, we treat the observed flux as a continuous field and develop an analytical model for 1D two-point flux statistics of metal ions in quasar spectra.
Our model makes use of three quantities to describe the two-point statistics: the column density distribution $f(N)$, an effective Doppler parameter $b_\mathrm{eff}$, and cloud-cloud clustering $\xi_{cc}$.
We decompose the correlation function contributions into one-absorber (1a) and two-absorber (2a) terms, where the 1a term captures the correlation of the doublet's shape with itself and the 2a term captures the clustering of two difference systems.
The most important difference with the halo model is that the 1a term does not correspond to metal (matter) clustering, but to a tightly bound single system that manifests Doppler broadening in its absorption profile.
Since prominent features are due to doublet transitions, the 1D flux field correlation function (power spectrum) shows a peak (oscillations) at the doublet separation scale even at the one-absorber level.

We apply our model to power spectra measurements from high-resolution spectra and DESI early data.
For this exploratory application, we focus on \ion{C}{IV} values and marginalize over \ion{Si}{IV} and \ion{Mg}{II}.
We use high-resolution \poned as is, and decompose DESI data into two samples: with and without DLA and sub-DLA sight-lines.
All three data sets show a clear increase in the amplitude of \ion{C}{IV} with decreasing redshift, although this trend could partially be due to the growth of cloud-cloud clustering.
The results from high-resolution spectra and non-DLA DESI sight-lines agree well between $2.0<z<3.0$.
However, the abundance results using all DESI spectra are noticably larger.
Significant difficulties for our model arise from fitting superposed oscillations in the power spectrum, the sensitivity to contaminants, and unmodelled large-scale power.
For example, it is unclear how well we can constrain the $b$ parameter (therefore temperature), so we refrain from making cosmological statements about the temperature of these systems.
Making robust claims also requires a careful study of systematics in the data and related numerical artifacts, as well as improved modelling of the large-scale correlations.
We plan to improve our model and numerical approach to apply for the next set of DESI spectra.

\section*{Acknowledgments}
\label{sec:acknow}
NGK thanks Matthew Pieri, Eric Armengaud and Ignasi Pérez-Ràfols for valuable comments and discussions on the draft.

This research used resources of the National Energy Research Scientific Computing Center, which is supported by the Office of Science of the U.S. Department of Energy under Contract No. DE-AC02-05CH11231.
Our quadratic estimator is written in \textsc{c++}.
It depends on \textsc{cblas} and \textsc{lapacke} routines for matrix/vector operations, \textsc{GSL}\footnote{\url{https://www.gnu.org/software/gsl}} for certain scientific calculations \citep{GSL}, \textsc{FFTW3}\footnote{\url{https://fftw.org}} for deconvolution when needed \citep{FFTW05}; and uses the Message Passing Interface (MPI) standard\footnote{\url{https://www.mpi-forum.org}}$^,$\footnote{\url{https://www.mpich.org}}$^,$\footnote{\url{https://www.open-mpi.org}} to parallelize tasks.
We use the following staple software in \textsc{python} analysis: \textsc{astropy}\footnote{\url{https://www.astropy.org}}
a community-developed core Python package for Astronomy \citep{astropy:2013, astropy:2018, astropy:2022},
\textsc{numpy}\footnote{\url{https://numpy.org}}
an open source project aiming to enable numerical computing with \textsc{python} \citep{numpy},
\textsc{scipy}\footnote{\url{https://scipy.org}}
an open source project with algorithms for scientific computing,
\textsc{numba}\footnote{\url{https://numba.pydata.org}}
an open source just-in-time (JIT) compiler that translates a subset of \textsc{python} and \textsc{numpy} code into fast machine code,
\textsc{mpi4py}\footnote{\url{https://mpi4py.readthedocs.io}}
which provides \textsc{python} bindings for the MPI standard \citep{mpi4py}.
Finally, we make plots using
\textsc{matplotlib}\footnote{\url{https://matplotlib.org}}
a comprehensive library for creating static, animated, and interactive visualizations in \textsc{python}
\citep{matplotlib}.

This research is supported by the Director, Office of Science, Office of High Energy Physics of the U.S. Department of Energy under Contract No. DE–AC02–05CH11231, and by the National Energy Research Scientific Computing Center, a DOE Office of Science User Facility under the same contract; additional support for DESI is provided by the U.S. National Science Foundation, Division of Astronomical Sciences under Contract No. AST-0950945 to the NSF’s National Optical-Infrared Astronomy Research Laboratory; the Science and Technologies Facilities Council of the United Kingdom; the Gordon and Betty Moore Foundation; the Heising-Simons Foundation; the French Alternative Energies and Atomic Energy Commission (CEA); the National Council of Science and Technology of Mexico (CONACYT); the Ministry of Science and Innovation of Spain (MICINN), and by the DESI Member Institutions: \url{https://www.desi.lbl.gov/collaborating-institutions}.

The authors are honored to be permitted to conduct scientific research on Iolkam Du’ag (Kitt Peak), a mountain with particular significance to the Tohono O’odham Nation.

\section*{Data availability}
All data points shown in figures are available in simple text files on the following website: \url{https://doi.org/10.5281/zenodo.7548373}.
Some underlying DESI spectra will be publicly available as early data release (EDR) in 2023.
We added two months of main surveys to improve our statistics.
The main survey spectra will be made publicly available as part of Year 1 data release in the future.

\bibliographystyle{mnras}
\bibliography{ion-abundance-references.bib}
\bsp	
\label{lastpage}
\end{document}